\documentclass[pdflatex,sn-basic]{sn-jnl}

\usepackage{graphicx}%
\usepackage{multirow}%
\usepackage{amsmath,amssymb,amsfonts}%
\usepackage{amsthm}%
\usepackage{mathrsfs}%
\usepackage[title]{appendix}%
\usepackage{xcolor}%
\usepackage{textcomp}%
\usepackage{manyfoot}%
\usepackage{booktabs}%
\usepackage{algorithm}%
\usepackage{algorithmicx}%
\usepackage{algpseudocode}%
\usepackage{listings}%

\theoremstyle{thmstyleone}%
\theoremstyle{thmstyletwo}%

\theoremstyle{thmstylethree}%

\begin{document}

\title{Gravitational Wave Distance Estimation Using Intrinsic Signal Properties: Dark Sirens as Distance Indicators}

\author[1]{\fnm{Trisha} \sur{V}}\email{trisha.v@phy.christuniversity.in}
\equalcont{These authors contributed equally to this work.}

\author[1]{\fnm{Rakesh} \sur{V}}\email{rakesh.v@phy.christuniversity.in}
\equalcont{These authors contributed equally to this work.}

\author*[1]{\fnm{Arun} \sur{Kenath}}\email{kenath.arun@hristuniversity.in}

\affil*[1]{\orgdiv{Department of Physics and Electronics}, \orgname{Christ University}, \orgaddress{ \city{Bengaluru}, \postcode{560029}, \state{Karnataka}, \country{India}}}

\abstract{Gravitational Waves (GWs) provide a powerful means for cosmological distance estimation, circumventing the systematic uncertainties associated with traditional electromagnetic (EM) indicators. This work presents a model for estimating distances to binary black hole (BBH) mergers using only GW data, independent of EM counterparts or galaxy catalogs. By utilizing the intrinsic properties of the GW signal, specifically the strain amplitude and merger frequency, our model offers a computationally efficient preliminary distance estimation approach that could complements existing Bayesian parameter estimation pipelines.

In this work, we examine a simplified analytical expression for the GW luminosity distance derived from General Relativity (GR), based on the leading-order quadrupole approximation. Without incorporating post-Newtonian (PN) or numerical relativity (NR) corrections, or modeling spin, eccentricity, or inclination, we test how closely this expression can reproduce distances reported by full Bayesian inference pipelines. We apply our model to 87 events from the LIGO-Virgo-Kagra (LVK) Gravitational Wave Transient Catalogues (GWTC), computing distances for these sources. Our results demonstrate consistent agreement with GWTC-reported distances, further supported by graphical comparisons that highlight the model’s performance across multiple events.}


\maketitle
\section{Gravitational Waves as Standard Sirens for Cosmological Distance Measurement}
\label{sec:intro}
Gravitational Waves, first predicted by Einstein in 1916, provide a new method for measuring cosmological distances. The ability to directly measure distances using GW signals, referred to as "standard sirens", was initially proposed by \citep{schutz1986}. Unlike electromagnetic-based distance indicators, such as Type Ia supernovae or Cepheid variables, which rely on the cosmic distance ladder and are subject to various sources of systematic error, GWs offer a more direct and independent method of measuring distances \citep{schutz1986}. The amplitude of the GW signal encodes the distance to the source, while the waveform's shape provides information about the mass and spin of the binary system, making GWs ideal for distance measurement.

The potential of standard sirens was first realized with the detection of GWs from the BBH merger GW150914 during LIGO-Virgo-Kagra (LVK) O1 run \citep{abbott2016}. This detection demonstrated the viability of GW astronomy but was limited to events without an associated EM signal, making it impossible to directly measure the redshift of the source. The first demonstration of the full potential of standard sirens occurred in 2017 with the observation of GW170817, a neutron star merger that was accompanied by a gamma-ray burst and optical counterpart. This multi-messenger event allowed astronomers to measure both the GW signal and the redshift of the host galaxy, enabling an independent estimation of the Hubble constant \citep{abbott2017}.

The success of GW170817 highlighted the power of standard sirens when paired with EM counterparts, but it also underscored the limitations of such an approach. Most GW events, particularly those involving black hole mergers, do not produce detectable EM signals. This has led to significant interest in developing methods for using GW signals alone to measure distances without the need for EM counterparts. Such events, referred to as dark sirens, hold the key to unlocking the full potential of GW astronomy for cosmological distance measurements.

\subsection{Dark Sirens and Non-EM Distance Estimation}

In cases where no EM counterpart is observed, distance measurement becomes significantly more challenging. Traditional standard siren analyses rely on associating a GW event with a host galaxy to obtain a redshift, a step that is infeasible for the majority of detected BBH mergers \citep{fishbach2019}. As a result, the majority of GW detections cannot be used for standard siren cosmology under traditional methods.

Recent efforts have explored alternative approaches for estimating distances using dark sirens. One method statistically associates GW events with galaxy catalogues, matching the event’s localization region to known galaxies in the area \citep{fishbach2019}. While this provides a viable approach, it introduces significant uncertainties due to the large sky localization areas of GW detections, which can span hundreds of square degrees.

To address these limitations, researchers have focused on using the intrinsic properties of the GW signal itself, such as strain amplitude and merger frequency, to infer distances. \citep{holz2005} demonstrated that GWs could serve as self-contained standard sirens, using the amplitude of the GW signal as a distance indicator while deriving binary parameters from waveform analysis. Despite these advances, most existing methods assume the presence of an EM counterpart, limiting their general applicability. 

\subsection{Gravitational Wave Astronomy and LVK's Observational Runs} GW astronomy has transformed the way astrophysical phenomena are studied since the first detection of a BBH  merger by the LVK collaboration in 2015 \citep{abbott2016}. The subsequent observation runs—O1, O2, and O3—have provided a wealth of data, enabling the identification of numerous GW events, such as binary neutron star (BNS) and BBH mergers. The LVK network of ground-based interferometric detectors has completed three observing runs so far. These have provided over 180 GW detections \citep{abbott2021}, \citep{abbott2020}, \citep{zackay2019b}, \citep{abbott2019a},  \citep{nitz2021a}, \citep{Nitz2021b}. These observations have allowed researchers to extract important information regarding the masses, spins, and distances of the systems that produce these signals. The increased sensitivity during each of these runs has expanded the GW catalogue, especially after the O3 run, which detected 56 new events \citep{abbott2021}, making the total number of confirmed GW detections 90 by the end of O3. The data provided by LVK's O1, O2, and O3 runs are pivotal in studying compact objects, as they allow for detailed parameter estimation and a deeper understanding of BBH systems. One of the most important parameters that can be extracted from these detections is the distance to the GW source, which is essential to constrain cosmological models and improve our understanding of the expansion of the universe. However, unlike EM observations, which often have reliable distance indicators, measuring distances with GWs requires careful modelling of the waveform and its associated parameters. 

\subsection{ Gravitational Waves as Standard Sirens } GWs can be used as ``standard sirens'', a concept first proposed by \citep{schutz1986} and later refined by \citep{holz2005}. Similar to how astronomers use ``standard candles'' such as Type Ia supernovae to measure cosmological distances, standard sirens allow for the direct estimation of distances from GW signals alone. The advantage of GWs over EM observations is that the distance measurement can be obtained directly from the waveform, independent of complex astrophysical assumptions about the source's intrinsic brightness or characteristics. Initial GW detections that provided accurate distance measurements often relied on the presence of an EM counterpart, as seen in the binary neutron star merger GW170817 \citep{abbott2017}. This event, which was followed by a kilonova and a gamma-ray burst, enabled precise distance estimation by combining both GW and EM data. However, for BBH systems, no such EM counterparts are expected, which poses a significant challenge for distance measurement. Several studies have aimed to address this gap by developing methods to estimate distances without the need for EM counterparts. \citep{chen2018} explored the feasibility of using GWs alone for distance estimation, with particular emphasis on BBH mergers. Although their work has demonstrated the potential for standard sirens to measure distances with some accuracy, significant uncertainties remain, particularly due to the degeneracy between the orientation of the binary system and the inferred distance. This study seeks to improve upon these methods by using the strain and merger frequencies from LVK’s O1, O2, and O3 data to estimate distances without relying on EM observations. By focussing on these two key parameters, it is possible to reduce the uncertainties that have traditionally plagued distance measurements in BBH systems.

\subsection{Standard Siren Methods Using Neutron Stars}

While BNS mergers like GW170817 serve as bright standard sirens, meaning they have EM counterparts for direct redshift measurement, neutron star mergers also provide alternative GW-only methods. One promising approach is the measurement of tidal deformations, which encode information about the neutron star equation of state (EoS). The imprint of tidal interactions on the GW waveform allows for constraints on the redshift-independent properties of the system, which, when combined with population statistics, can improve distance estimates \citep{abbott2017}.  

While neutron star standard sirens have clear advantages, they are rare events, with BNS merger detections significantly outnumbered by BBH mergers. Since BBH mergers do not experience tidal deformations or produce EM counterparts, alternative GW-only distance estimation methods are required.

\subsection{Waveform Models in Gravitational Wave Analysis: IMRPhenomD} The accuracy of distance estimation from GW data is contingent upon the waveform model used to interpret the signal. The \textit{IMRPhenomD} model is widely regarded as one of the most reliable waveform models for studying the inspiral, merger, and ringdown phases of BBH systems \citep{husa2016, khan2016}. This model provides an effective means of parameter estimation by combining aspects of both analytical and numerical relativity to describe the GW signal generated during the coalescence of two black holes. IMRPhenomD is particularly well-suited for analysing data from BBH systems, as it accounts for non-precessing spins and covers a wide range of mass ratios. The model has been used extensively in the analysis of LVK data, helping to extract critical parameters such as the strain and frequency of the GWs. By applying the IMRPhenomD model to the GW signals from the O1, O2, and O3 runs, precise estimates of the distance to the BBH systems can be obtained. In this study, the IMRPhenomD model is employed to extract strain and merger frequencies from the GWTC data. These parameters are then used in a novel model to estimate distances, providing a method for studying BBH systems without requiring EM observations. This approach not only improves the precision of distance measurements but also enables better constraints on other key parameters, such as the masses and spins of the black holes involved in the merger. 

\subsection{Current Challenges in Distance Measurement without Electromagnetic Counterparts} Despite the progress made in measuring distances using GWs, several challenges persist, particularly when EM counterparts are absent. One of the most significant challenges is the degeneracy between the distance to the source and the inclination angle of the binary system \citep{abbott2017}. This degeneracy arises because the GW strain, which is used to estimate distance, is affected by both the distance to the source and the orientation of the binary system relative to the observer. As a result, distinguishing between a distant, face-on system and a closer, edge-on system can be difficult without additional information. Another challenge arises from the inherent uncertainties in waveform modelling. While the IMRPhenomD model is highly successful in describing a wide range of BBH systems, it still relies on certain approximations, particularly in modelling the merger and ringdown phases. These approximations can introduce uncertainties into the estimation of key parameters, including distance \citep{khan2016}. This study aims to address these challenges by focusing on the strain and merger frequency parameters, which provide a more robust basis for distance estimation. By refining these measurements, it is possible to reduce the uncertainties that arise from the degeneracy between distance and inclination, as well as from waveform modelling errors. This method offers a promising alternative to existing techniques and has the potential to improve the precision of future GW observations. 

\subsection{Preliminary Measurements and their Role in Constraining BBH Parameters} Preliminary studies have demonstrated that accurate distance estimation using GWs alone can provide important constraints on other BBH parameters, such as mass and spin \citep{abbott2020}. For instance, the LVK collaborations have shown that by focussing on the GW strain and frequency evolution, it is possible to reduce the uncertainties in the mass and spin measurements of the black holes involved in the merger. This has important implications for understanding the population of black holes in the universe, as well as for testing models of black hole formation and evolution. In this study, the focus is on refining the strain and merger frequency parameters to improve the accuracy of distance measurements. By applying the IMRPhenomD model to the data from GWTC’s O1, O2, and O3 runs, precise estimates of the distances to BBH systems are obtained, allowing for better constraints on other parameters. This approach not only enhances the precision of distance measurements but also contributes to the broader understanding of black hole physics and cosmology. 

While previous works have used GWs to estimate distances through Bayesian parameter estimation frameworks, we adopt a different approach. Rather than developing a new model, we revisit the analytic expressions derived from GR for binary inspiral systems in the quadrupole approximation.

Our goal is to test how closely this basic theoretical expression—excluding higher-order PN or NR corrections and without incorporating orbital inclination or spin can approximate luminosity distances when evaluated with empirical parameters from GWTC events. This allows us to validate the expression’s applicability as a quick, model-light estimator and a pedagogical cross-check on more complex inference pipelines.

\section{Gravitational Wave Analysis for Cosmological Distance Estimation}
\label{sec:methodology}
This section describes the methodology used to estimate the distance to binary black hole (BBH) systems using gravitational wave (GW) data. The approach relies solely on GW signals, without requiring electromagnetic (EM) counterparts. The IMRPhenomD waveform model is used to extract key parameters, such as the strain amplitude and merger frequency, from the GW signals detected by the LVK detectors. These parameters are then used to compute the distance to the BBH systems based on a purely GW-based approach.

The methodology is grounded in the analytic expressions for gravitational wave luminosity and orbital decay derived directly from GR. We do not attempt to reconstruct full waveforms or perform parameter estimation; instead, we explore how well the GR-predicted strain–distance relation holds when applied to observed binary black hole mergers. 

\subsection{Binary Black Hole Systems and Gravitational Wave Emission}
\label{subsec:BBH}

In this study, we consider BBH systems in which the two black holes orbit each other in circular orbits around their centre of mass. The BBHs are assumed to be face on and the separation between the black holes and their velocities is key to determining the GW emission \citep{Misner1973}.

We consider BBH systems where two black holes orbit each other in nearly circular orbits. The separation between the black holes and their velocities determine the strength and frequency of the emitted GW signal. Under the assumption of a non-precessing, quasi-circular binary, the GW frequency $f_{\text{GW}}$ is related to the orbital frequency $f_o$ by:

\begin{itemize}
    \item The GW frequency is twice that of the orbital frequency $f_o$:
    \begin{equation}
    f = 2f_o
    \end{equation}
  \item The orbital separation at the time of merger \( r_0 \) is computed using the relation:
    \begin{equation}
   r_0 = \left( \frac{G M}{(2\pi f_{\text{o}})^2} \right)^{1/3}
    \end{equation}
    where $G$ is the gravitational constant, and $r$ is the separation between black holes.
\end{itemize}

\subsection{Gravitational Wave Luminosity and Energy Loss}

GWs carry energy away from the binary system, causing the orbit to shrink over time \citep{michelle2007}. The GW luminosity $L_{\text{GW}}$ for a binary system is expressed as:
\begin{equation*}
    L_{GW} = -\frac{G}{5c^5} \langle\dddot{Q}_{ij}  \dddot{Q}^{ij} \rangle
\end{equation*}
\begin{equation}
L_{\text{GW}} = \frac{-32 G^4 (M_1 M_2)^2 M}{5c^5 r^5}
\end{equation}
where $c$ is the speed of light and ${Q}_{ij}$ is the reduced quadrupole moment. This energy loss causes the orbit to decay over time. The rate of orbital decay is:
\begin{equation}
\frac{dr}{dt} = -\frac{64G^3 M_1 M_2 M}{5c^5 r^3}
\end{equation}

The energy loss leads to a merger phase after a finite time, known as the merger time $t_{\text{merger}}$, the time taken for the BBH to coalesce from an initial orbital separation $r_0$ , which is:
\begin{equation}
t_{\text{merger}} = \frac{r_0^4}{4 \kappa}, \quad \kappa = \frac{64G^3 M_1 M_2 M}{5c^5}
\end{equation}

\subsection{Chirp Mass and Frequency Evolution}

The evolution of the GW frequency is governed by the chirp mass $M_c$, which is a combination of the component masses:
\begin{equation}
M_c = \frac{(M_1 M_2)^{3/5}}{(M_1 + M_2)^{1/5}}
\end{equation}
The rate of change of the GW frequency $\dot{f}$ is given by:
\begin{equation}
\dot{f} = \frac{96}{5} \pi^{8/3} \left( \frac{G M_c}{c^3} \right)^{5/3} f^{11/3}
\end{equation}

\subsection{Gravitational Wave Strain and Distance to the Source}

The strain $h_o$ of the GW as detected on Earth depends on the chirp mass $M_c$, the distance $D$ to the source, and the frequency of the GW. The strain is given by:

\begin{equation}
h_o = \frac{4 G M_c}{D c^2} \left( \frac{G \pi M_c f}{c^3} \right)^{2/3}
\end{equation}
\raggedright
From this equation, the distance $D$ to the source can be calculated as:
\begin{equation}
D = \frac{5 c \dot{f}}{24 \pi^2 h_o f^3}
\end{equation}
or equivalently:
\begin{equation}
D = \frac{5c}{64 \pi^2 h_o f^2 t_{\text{merger}}}
\end{equation}
where
    $h_o$ is the GW strain,
    $f$ is the GW frequency and
    $t_{\text{merger}}$ is the duration of the GW signal from the inspiral to the merger.

\subsection{Application of the Model to GWTC Data}

This methodology was applied to data from the LVK O1, O2 and O3 observation runs. These runs provided a wealth of GW data, primarily from BBH mergers, which are catalogued in the GWTC. The goal of this study is to estimate the distance to BBH systems using only GW data, without relying on the EM counterparts.

\subsubsection{Parameter Extraction}

Additional parameters necessary for waveform modelling, including the mass ratio $q = M_2 / M_1$ $(M_2 < M_1$) and the spin parameters ($\chi_1$ and $\chi_2$) of the black holes, were extracted from the \textit{parameter estimation (PE)} h5 files available on the GWOSC website. These files provide posterior distributions of the physical parameters for each detected event.

The mass ratio and spin parameters, along with the total masses, were used to model the gravitational waveform for each BBH event using the IMRPhenomD model.

\subsubsection{Waveform Modelling using IMRPhenomD}

The IMRPhenomD waveform model was employed to generate the gravitational waveforms for the BBH events in the GWTC catalogues. IMRPhenomD is a phenomenological model that combines numerical relativity and post-Newtonian approximations to describe the full waveform for BBH systems.

The following key inputs were used for waveform modelling:
\begin{itemize}
    \item \textbf{Total Mass} ($M_t$): Available in the GWOSC catalogue.
    \item \textbf{Mass Ratio} ($q$): Affects the waveform amplitude and phase evolution.
    \item \textbf{Spin Parameters} ($\chi_1$ and $\chi_2$): The black hole spins influence the waveform morphology.
\end{itemize}

The IMRPhenomD model provided the GW strain $h_o$ and the merger frequency $f$ for each event, which were then used to estimate the distance to the source.

\subsubsection{Component Mass and Distance Estimation Methodology}

For each GW event, the component masses ($M_1, M_2$) are derived using the total mass ($M_t$) and the mass ratio ($q = M_2/M_1$) obtained from the posterior distributions available in the GWTC catalogues. To account for uncertainties in mass measurements and their impact on distance estimation, we compute distances for three distinct sets of component masses:
\begin{itemize}
    \item Maximum mass asymmetry scenario: $M_1$ is set to its highest possible value, while $M_2$ is set to its lowest possible value within the reported mass range. This represents the extreme case where the binary has the most unequal mass distribution.
    \item Minimum mass asymmetry scenario: $M_1$ is set to its lowest possible value, while $M_2$ is set to its highest possible value within the mass range .This corresponds to the scenario where the mass distribution is reversed but still within the allowable range.
    \item Median mass values: Both $M_1$ and $M_2$ are set to their median values from the GWTC posterior distributions. This represents the most likely component mass configuration based on the Bayesian parameter estimation results.

\end{itemize}

By evaluating distances across these three mass configurations, we ensure that the computed distances capture the full range of mass uncertainty effects, providing a more comprehensive and reliable estimation. The final distance uncertainty $\sigma_D$ is determined by propagating the uncertainties in mass, strain amplitude, and frequency through the distance calculation equation

This approach is purely GW-based and does not rely on EM observations, making it ideal for BBH systems, which typically lack detectable EM counterparts.

\subsection{Data Sources and Analysis Procedures}

The GW distance estimation model was validated using data from the GWOSC, specifically the GWTC. This catalogue includes all GW events detected by the LVK collaborations during the O1, O2, and O3 observational runs. For this study, we focused exclusively on BBH  mergers, as these events provide the necessary data for validating the model.
The primary parameters obtained from the GWTC catalogue  included:
\begin{itemize}
    \item The total mass and mass ratio of the binary systems ($m_t$ and $q$),
    \item The signal-to-noise ratio and strain data for the GW events,
    \item GWTC-reported luminosity distances for each event.
\end{itemize}

\subsection{Waveform Modelling with IMRPhenomD}

To calculate the GW strain ($h_0$) and frequency ($f$), we employed the IMRPhenomD waveform model. This model provides an accurate representation of the GW signals emitted during the inspiral, merger, and ringdown phases of binary coalescence events. The mass values from the GWOSC catalogue are input into the IMRPhenomD model to generate the required strain and frequency values.                               

This model does not aim to derive the full waveform evolution from first principles or use it in a parameter estimation framework. Instead, we validate the analytical expression for luminosity distance (Eq. 10) by utilizing selected posterior samples from the GWTC catalog. Specifically, we extract the component masses and the reported luminosity distance from the GWTC posteriors and use these inputs in the IMRPhenomD waveform model. This procedure allows us to compute the waveform and extract key phenomenological quantities, including the merger time and merger frequency. These intrinsic quantities are then inserted into Eq. (10), which we test for consistency against LIGO-Virgo-KAGRA observations.
    
Our approach is model-informed but does not rely on detailed post-Newtonian or numerical relativity expansions in deriving Eq. (10). Rather, the IMRPhenomD waveform family is employed as a practical tool to extract the characteristic signal timescales and frequencies, enabling us to assess the validity of Eq. (10) across a representative sample of empirically observed binary black hole mergers.

\subsection{Distance Calculation}
 The distance to each GW event was calculated using Equation 10.

\subsubsection{Event Selection}
We used all BBH events from the GWTC confident detection catalogue , ensuring high detection confidence and complete mass data for each event. This allowed us to apply the mathematical model to a broad range of GW detections for BBH mergers.

\subsubsection{Statistical and Error Analysis}
A statistical analysis was conducted to assess the performance of our model across the full set of GW events. The average deviation between the calculated and GWTC-reported distances were computed to evaluate consistency. An error analysis was performed to account for uncertainties in the binary mass measurements and strain values, assessing how these uncertainties propagate through the distance estimation process.

The total uncertainty in the estimated distance ($\sigma_D$) is determined by propagating the uncertainties in component masses and strain amplitude. Using standard error propagation, the total variance in distance can be approximated as:
\begin{equation}
\sigma_D^2 = \left( \frac{\partial D}{\partial M_1} \sigma_{M_1} \right)^2 + \left( \frac{\partial D}{\partial M_2} \sigma_{M_2} \right)^2 + \left( \frac{\partial D}{\partial h_0} \sigma_{h_0} \right)^2.
\end{equation}
Here, $\sigma_{M_1}$ and $\sigma_{M_2}$ represent the uncertainties in component masses, and $\sigma_{h_0}$ is the uncertainty in the GW strain amplitude.

\section{Results and Discussion}

\begin{table}[!]
\footnotesize
\begin{tabular}{@{}llllllll}
    \textbf{Event} & $m_1$ $(M_\odot)$ & $m_2$ $(M_\odot)$ & \textbf{$f_{GW}$(Hz)} & \textbf{$h_0$ (max strain)} & \textbf{$d_{Model}$ [Mpc]} & \textbf{$d_{GWTC}$ [Mpc]} \\ \hline
        ~ & 35.60 & 30.60 & 161.04 & $1.79 \times 10^{-21}$ & ~ & ~ \\  
        \textbf{GW150914} & 40.3 & 26.2 & 186.26 & $1.72 \times 10^{-21}$ & $540^{+100}_{-60}$ & $440^{+150}_{-170}$ \\  
        ~ & 33.6 & 32.5 & 160.78 & $1.79 \times 10^{-21}$ & ~ & ~ \\  
        \hline
        ~ & 23.2 & 13.6 & 298.36 & $3.78 \times 10^{-22}$ & ~ & ~ \\  
        \textbf{GW151012} & 38.1 & 8.8 & 220.24 & $3.03 \times 10^{-22}$ & $1130^{+210}_{-120}$ & $1080^{+550}_{-490}$ \\  
        ~ & 17.7 & 17.7 & 345.92 & $3.92 \times 10^{-22}$ & ~ & ~ \\  
        \hline
        ~ & 13.7 & 7.7 & 561.17 & $5.21 \times 10^{-22}$ & ~ & ~ \\  
        \textbf{GW151226} & 22.5 & 5.2 & 398.78 & $4.31 \times 10^{-22}$ & $590^{+120}_{-70}$ & $450^{+180}_{-190}$ \\  
        ~ & 10.5 & 9.9 & 682.11 & $5.42 \times 10^{-22}$ & ~ & ~ \\  
        \hline
        ~ & 30.8 & 20 & 252.75 & $5.83 \times 10^{-22}$ & ~ & ~ \\  
        \textbf{GW170104} & 38.1 & 15.4 & 212.24 & $5.19 \times 10^{-22}$ & $1240^{+160}_{-140}$ & $990^{+440}_{-430}$ \\  
        ~ & 25.2 & 24.9 & 219.55 & $6.05 \times 10^{-22}$ & ~ & ~ \\  
        \hline
        ~ & 11 & 7.6 & 635.17 & $6.70 \times 10^{-22}$ & ~ & ~ \\  
        \textbf{GW170608} & 16.5 & 5.4 & 560.25 & $5.91 \times 10^{-22}$ & $420^{+60}_{-70}$ & $320^{+120}_{-110}$ \\  
        ~ & 9.3 & 9 & 591.09 & $6.83 \times 10^{-22}$ & ~ & ~ \\  
        \hline
        ~ & 35 & 23.8 & 194.81 & $6.56 \times 10^{-22}$ & ~ & ~ \\  
        \textbf{GW170809} & 43.3 & 18.6 & 180.98 & $5.94 \times 10^{-22}$ & $1320^{+140}_{-140}$ & $1030^{+320}_{-390}$ \\  
        ~ & 29.1 & 28.9 & 202.24 & $6.74 \times 10^{-22}$ & ~ & ~ \\  
        \hline
        ~ & 30.6 & 25.2 & 206.75 & $1.10 \times 10^{-21}$ & ~ & ~ \\  
        \textbf{GW170814} & 36.2 & 21.2 & 211.31 & $1.06 \times 10^{-21}$ & $780^{+100}_{-90}$ & $600^{+150}_{-220}$ \\  
        ~ & 28 & 27.6 & 211.35 & $1.11 \times 10^{-21}$ & ~ & ~ \\  
        \hline
        ~ & 27.7 & 9 & 317.07 & $4.41 \times 10^{-22}$ & ~ & ~ \\  
        \textbf{GW190412} & 21.7 & 11 & 356.59 & $4.81 \times 10^{-22}$ & $950^{+130}_{-100}$ & $720^{+240}_{-220}$ \\  
        ~ & 33.7 & 7.6 & 291.90 & $3.95 \times 10^{-22}$ & ~ & ~ \\  
        \hline
        ~ & 23.2 & 12.5 & 289 & $2.64 \times 10^{-22}$ & ~ & ~ \\  
        \textbf{GW190512\_180714} & 17.6 & 16 & 357.51 & $2.75 \times 10^{-22}$ & $1810^{+280}_{-250}$ & $1460^{+510}_{-590}$ \\  
        ~ & 28.8 & 9.9 & 274.56 & $2.35 \times 10^{-22}$ & ~ & ~ \\  
        \hline 
         ~ & 65 & 47 & 86.27 & $1.04 \times 10^{-21}$ & ~ & ~ \\  
        \textbf{GW191109\_010717} & 76 & 34 & 80.85 & $8.76 \times 10^{-22}$ & $1420^{+190}_{-240}$ & $1290^{+1130}_{-650}$ \\  
        ~ & 62 & 54 & 92.21 & $1.10 \times 10^{-21}$ & ~ & ~ \\  
        \hline
 ~ & 10.7 & 6.7 & 668.56 & $2.48 \times 10^{-22}$ & ~ & ~ \\  
        \textbf{GW191129\_134029} & 14.8 & 5 & 576.95 & $2.20 \times 10^{-22}$ & $1020^{+110}_{-110}$ & $790^{+260}_{-330}$ \\  
        ~ & 8.6 & 8.2 & 691.36 & $2.54 \times 10^{-22}$ & ~ & ~ \\  
        \hline  
        ~ & 45.1 & 34.7 & 162.55 & $3.12 \times 10^{-22}$ & ~ & ~ \\  
        \textbf{GW191222\_033537} & 56 & 24.2 & 129.1 & $2.64 \times 10^{-22}$ & $3890^{+640}_{-650}$ & $3000^{+1700}$ \\  
        ~ & 44 & 37.1 & 148.11 & $3.24 \times 10^{-22}$ & ~ & ~ \\  
        \hline 
        ~ & 12.1 & 8.3 & 596.49 & $1.82 \times 10^{-22}$ & ~ & ~ \\  
        \textbf{GW191126\_115259} & 17.6 & 5.9 & 474.12 & $1.60 \times 10^{-22}$ & $1650^{+230}_{-280}$ & $1620^{+740}$ \\  
        ~ & 10.2 & 9.9 & 515.54 & $1.86 \times 10^{-22}$ & ~ & ~ \\  
        \hline
        ~ & 40 & 32.7 & 155.04 & $5.03 \times 10^{-22}$ & ~ & ~ \\  
        \textbf{GW200224\_222234} & 46.7 & 25.5 & 144.89 & $4.58 \times 10^{-22}$ & $2150^{+330}_{-310}$ & $1710^{+500}_{-650}$ \\  
        ~ & 37.5 & 35.5 & 166.1 & $5.10 \times 10^{-22}$ & ~ & ~ \\  
        \hline
        ~ & 87 & 61 & 59.42 & $2.74 \times 10^{-22}$ & ~ & ~ \\  
        \textbf{GW200220\_061928} & 127 & 36 & 61.4 & $2.26 \times 10^{-22}$ & $6740^{+820}_{-750}$ & $6000^{+4800}_{-3100}$ \\  
        ~ & 87 & 64 & 61.24 & $2.88 \times 10^{-22}$ & ~ & ~ \\ 
        
    \end{tabular}

    \caption{Selected list of sources selected from GWTC-1, GWTC-2 and GWTC-3. For each GW event, three different mass values ($m_1, m_2$) are listed. These values are obtained using the total mass of the BBH system, and the mass ratio $q$. The table also includes the merger frequency ($f$) and strain ($h_0$), which are used for distance estimation. The distance estimates from GWTC data is provided for reference.}
    \label{table-1}
\end{table}

This section presents the results of applying our GW only distance estimation model to the events listed in the Gravitational-Wave Transient catalogue s (GWTC-1, GWTC-2.1, and GWTC-3) and compares them with GWTC’s distance estimates for these same events. The results are analyzed to assess the performance of our model, identify trends, and highlight any discrepancies or agreements between the two methods.
\subsection{Summary of Distance Estimates}

Tables \ref{GWTC-3}, \ref{GWTC-2.1}, and \ref{GWTC-1}  present the distances calculated using our model alongside the GWTC distances for each event. Overall, our model shows good agreement with GWTC’s estimates, although significant differences are observed in certain events, particularly in high-mass black hole systems.

\subsubsection{Key Observations}

The distance calculated by our model for selected few of GW sources are displayed in Table \ref{table-1} along with the GWTC estimates. Some of the key observations for these GW sources are listed below:

\begin{itemize}
    \item \textbf{GW150914 (GWTC-1)}: One of the earliest and most famous detections, our model gives a distance estimate of $540^{+50}_{-10}$ Mpc, which is comparable to GWTC’s estimate of \(440^{+150}_{-170}\) Mpc. The difference of around 100 Mpc is small, showing that our method aligns well for this BBH event with moderate component masses (\(m_1 = 35.6^{+4.7}_{-3.1} M_{\odot}\), \(m_2 = 30.6^{+3.0}_{-4.4} M_{\odot}\)).
    
    \item \textbf{GW170104 (GWTC-1)}: For this event, GWTC reports a distance of \(990^{+440}_{-430}\) Mpc, while our model estimates a distance of $1240^{+40}_{-20}$ Mpc, resulting in a difference of 250 Mpc. The discrepancy can be attributed to the higher component masses of this system (\(m_1 = 30.3^{+7.3}_{-5.6} M_{\odot}\), \(m_2 = 20.0^{+4.9}_{-4.6} M_{\odot}\)) and the fact that higher mass systems tend to produce shorter-duration signals.
    
    \item \textbf{GW190512\_180714 (GWTC-2.1)}: This event involves lower-mass black holes (\(m_1 = 23.3^{+5.3}_{-5.8} M_{\odot}\), \(m_2 = 12.6 ^{+3.6}_{-2.5} M_{\odot}\)) and produced a distance estimate of $1810^{+100}_{-70}$ Mpc in our model, compared to GWTC’s \(1460^{+510}_{-590}\) Mpc. The difference of 350 Mpc shows the consistency of our model for lower-mass events, where the inspiral phase contributes significantly to the detected signal.
    
    \item \textbf{GW191129\_134029 (GWTC-3)}: Our model provides a distance of $1020^{+10}_{-10}$ Mpc, while GWTC estimates \(790^{+260}_{-330}\) Mpc. The difference of 230 Mpc is reasonable and falls within the expected range of variability for these types of events, which have moderate component masses (\(m_1 = 10.7^{+4.1}_{-2.1} M_{\odot}\), \(m_2 = 6.7^{+1.5}_{-1.7} M_{\odot}\)).
    
    \item \textbf{GW191222\_033537 (GWTC-3)}: This high-mass event (\(m_1 = 45.1^{+1+10.9}_{-8.0} M_{\odot}\), \(m_2 = 34.7^{+9.3}_{-10.5} M_{\odot}\)) has a distance estimate of $3890^{+260}_{-250}$ Mpc from our model compared to GWTC’s \(3000^{+1700}_{-1700}\) Mpc. The difference of 890 Mpc is significant, which is consistent with our findings that larger discrepancies occur in higher-mass systems. The shorter-duration signals from high-mass mergers may explain these differences, as they are more sensitive to waveform modeling uncertainties.
    
    \item \textbf{GW191126\_115259 (GWTC-3)}: This event shows an excellent match between the two models. Our estimate of $1650^{+120}_{-60}$ Mpc is close to GWTC’s \(1620^{+740}_{-740}\) Mpc, with a difference of only 30 Mpc. This suggests that for moderately massive systems, both models produce consistent results, reinforcing the reliability of our method for such systems.
\end{itemize}

Our model provides a computationally efficient approach for estimating BBH distances, offering a preliminary first-pass estimate before full Bayesian parameter estimation. However, we do not claim that our method is more accurate than GWTC distances. Instead, our results indicate consistency within uncertainty limits, highlighting the feasibility of strain-based distance estimation.

\subsection{Analysis of the Differences}

The differences between the two models’ distance estimates can be attributed to several factors:

\subsubsection{Model Assumptions and Parameters}
\begin{itemize}
    \item \textbf{Waveform Models}: Both our model and GWTC’s distance estimates rely on waveform models, but the models may incorporate different assumptions about the mass ratio, spin, and other parameters. Our model focuses on using the GW strain and frequency at merger, while GWTC’s parameter estimation method incorporates a wider range of information, including priors on system inclination and spin.
    
    \item \textbf{Mass Ratios and Spins}: Events with large discrepancies between the two distance estimates often involve extreme mass ratios or high black hole spins. For example, in \textbf{GW200224\_222234 (GWTC-3)}, where the component masses are \(m_1 = 40.0^{+6.7}_{-4.5} M_{\odot}\) and \(m_2 = 32.7^{+4.8}_{-7.2} M_{\odot}\), our model estimates a distance of $2150^{+110}_{-90}$ Mpc, while GWTC estimates \(1710^{+500}_{-650}\) Mpc. The difference of 440 Mpc could be due to variations in how the mass ratio and spins are incorporated into the waveform models.
\end{itemize}

\subsection{Comparison of Model Distances with GWTC Data}
We evaluate the analytic GR expression for luminosity distance using component masses and distances obtained from the GWTC parameter estimation results. While these posterior samples incorporate waveform priors and Bayesian inference assumptions, we use them only as inputs to reconstruct the inspiral-merger timescale and frequency via the IMRPhenomD waveform model.

These reconstructed quantities are inserted into the analytic GR expression for distance. This is not a predictive model in itself, but a test of how closely the simplified expression approximates GWTC-reported distances under empirical conditions.

We present graphical comparisons of the model distances against those reported in the GWTC catalogues. These visualizations provide a clear representation of the agreement between our method and existing Bayesian estimates.
\begin{figure}[htbp]
    \centering
    \includegraphics[width=\textwidth]{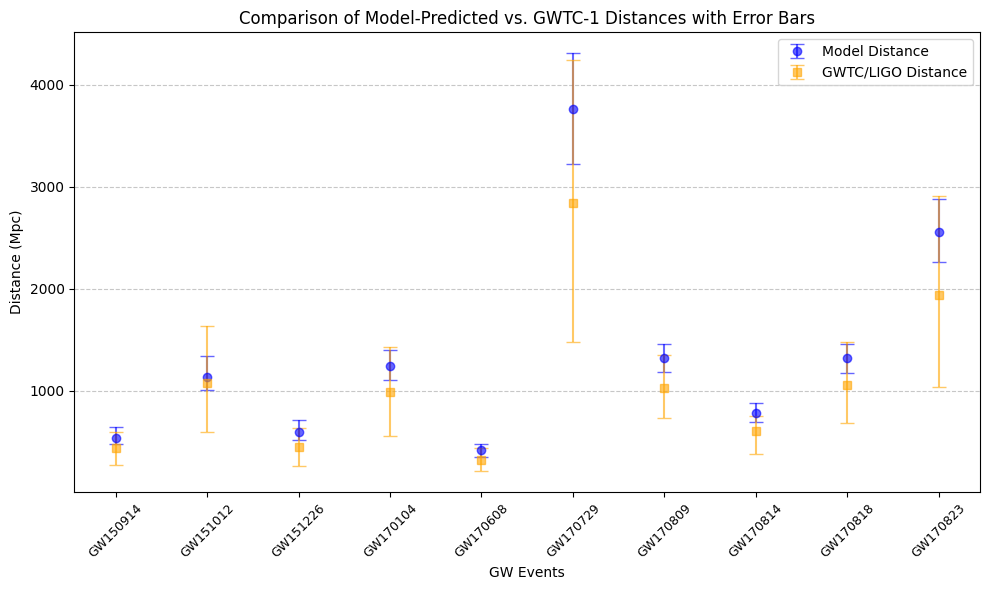}
    \caption{Comparison of Model vs. GWTC-1 Distances with Error Bars. The blue points represent the model distances, while the orange squares represent the GWTC/LIGO distances, with error bars indicating uncertainties in both estimates.}
    \label{fig:gwtc2}
\end{figure}

\begin{figure}[htbp]
    \centering
    \includegraphics[width=1\textwidth]{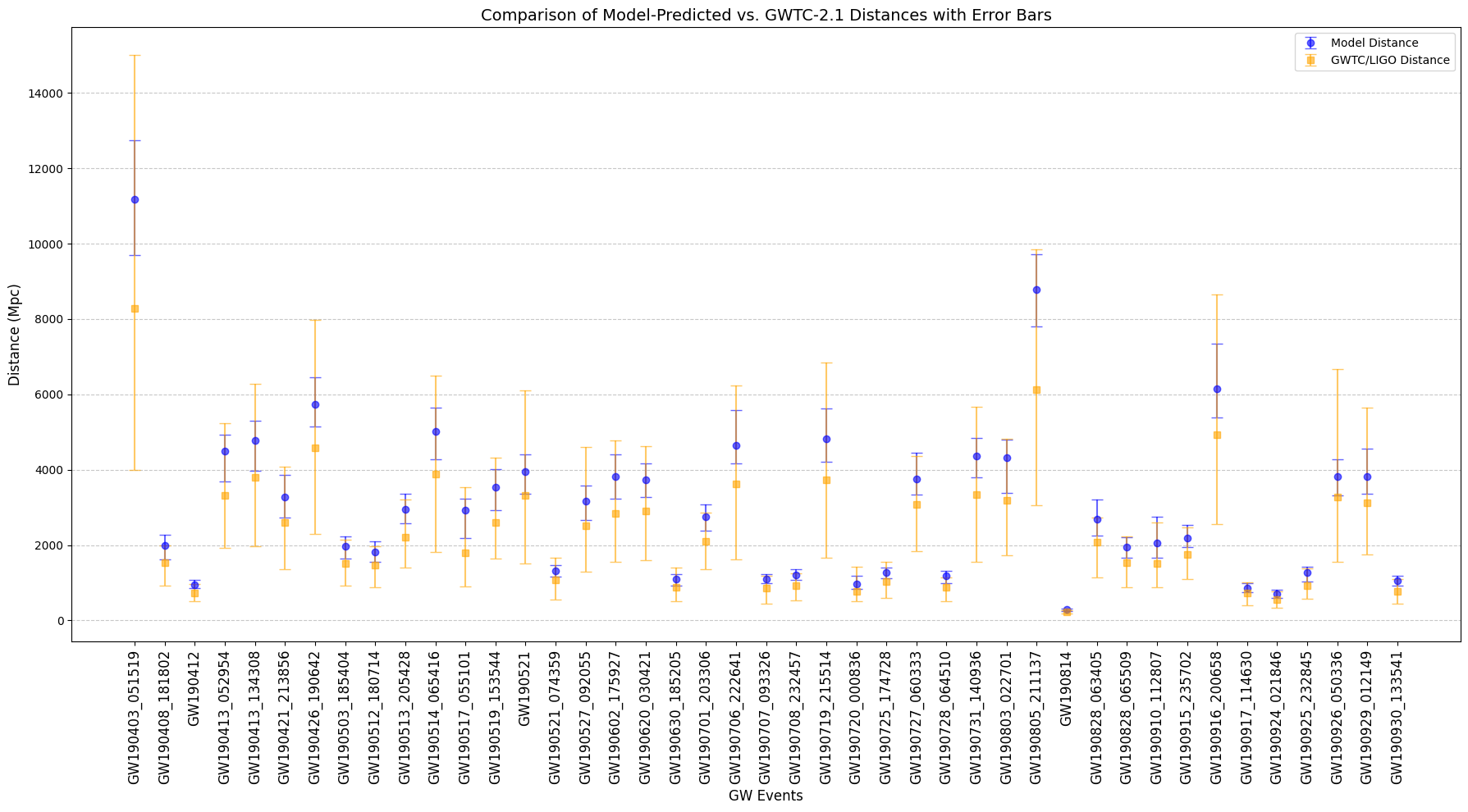}
    \caption{Comparison of Model vs. GWTC-2.1 Distances with Error Bars. The trends in this plot further illustrate the agreement between our method and Bayesian estimates.}
    \label{fig:gwtc3}
\end{figure}

\begin{figure}[htbp]
    \centering
    \includegraphics[width=1\textwidth]{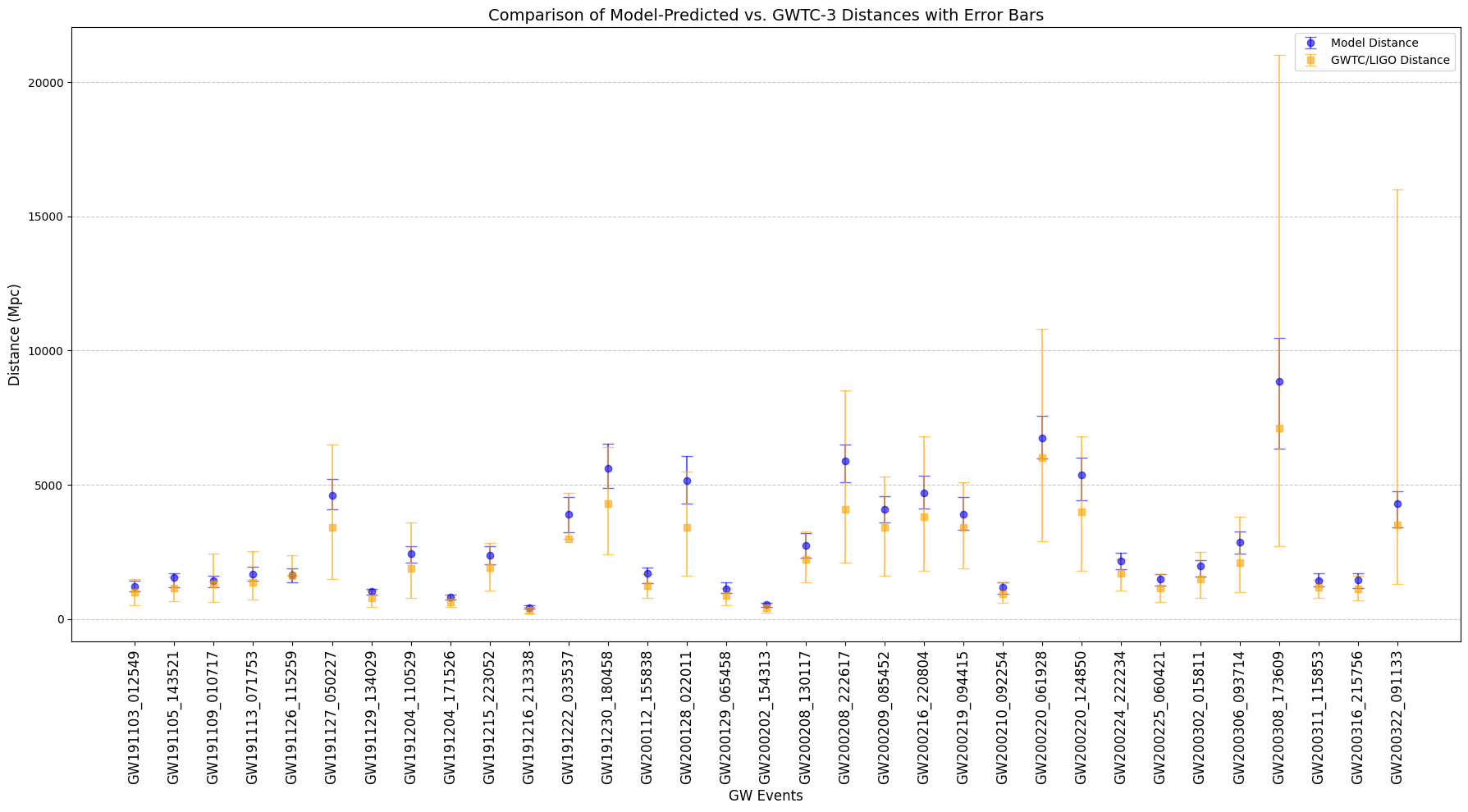}
    \caption{Comparison of Model vs. GWTC-3 Distances with Error Bars. The analysis of early events provides additional validation of the proposed method.}
    \label{fig:gwtc1}
\end{figure}

These plots highlight the overall consistency of our method with the GWTC results, while also revealing variations in individual cases. Notably, differences in distance estimates are observed in certain high-mass events, potentially due to waveform systematics or inclination angle effects. Additionally, the spread in error bars suggests that while our approach provides comparable results, further refinements in mass and inclination angle considerations could improve precision.

Our method derives strain amplitude and merger frequency from parameter estimation (PE) results available in the GWTC catalogues. Since PE incorporates waveform priors and Bayesian parameter inference, our model inherits some assumptions and uncertainties from the PE framework. This introduces potential inductive bias, as our method is not fully decoupled from the PE process. Future improvements should explore deriving strain amplitude directly from raw GW detector data to reduce dependence on PE-inferred parameters.

\begin{figure}[H]
    \centering
    \includegraphics[width=0.8\textwidth]{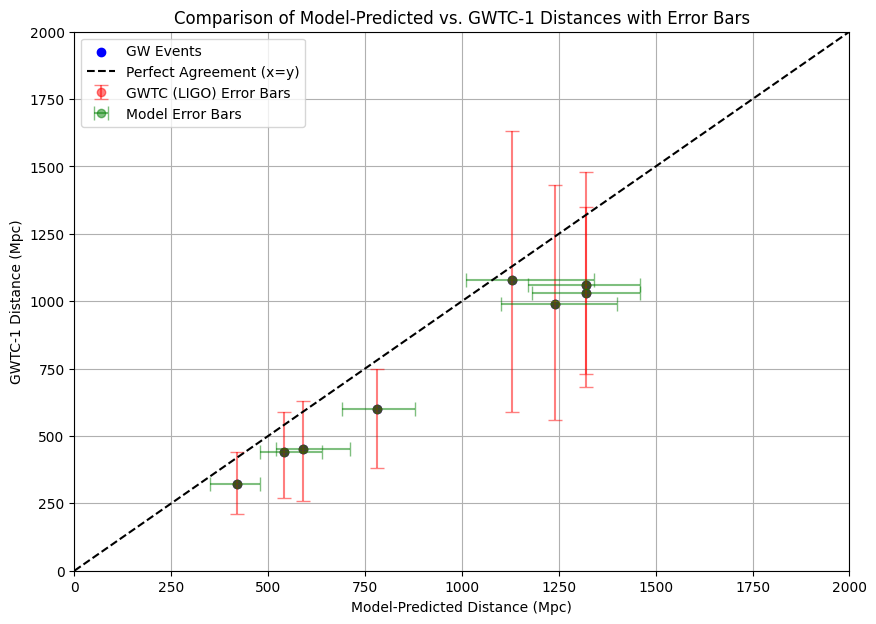}
    \caption{Comparison of model distances vs. GWTC-1 distances. The general alignment of points along the diagonal suggests consistency between the model and GWTC-1 distance estimates.}
    \label{fig:gwtc1-1}
\end{figure}

\begin{figure}[H]
    \centering
    \includegraphics[width=0.8\textwidth]{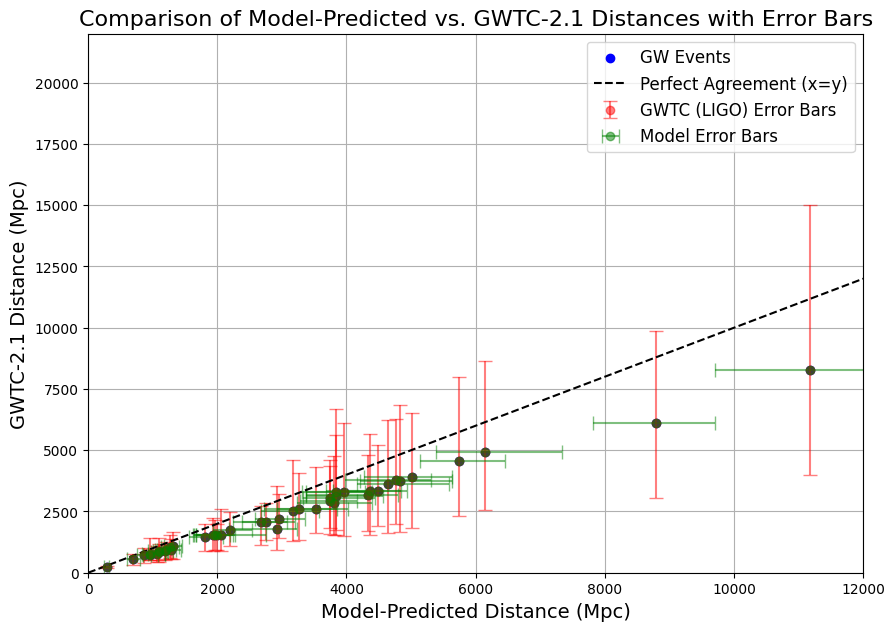}
    \caption{Comparison of model distances vs. GWTC-2.1 distances. While the model shows good agreement for lower distances, deviations increase at higher values, emphasizing the need for further refinement.}
    \label{fig:gwtc2-1}
\end{figure}

\begin{figure}[H]
    \centering
    \includegraphics[width=0.8\textwidth]{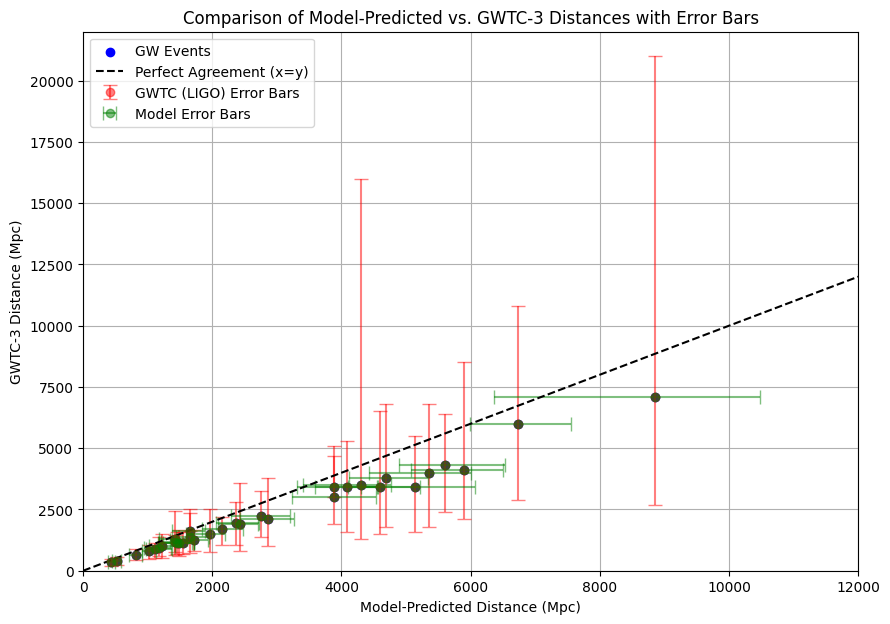}
    \caption{Comparison of model distances vs. GWTC-3 distances. The black dashed line represents perfect agreement (x=y). Green error bars correspond to model uncertainties, while red error bars represent GWTC-reported uncertainties.}
    \label{fig:gwtc3-1}
\end{figure}

The plots presented above compare the model distances with the gravitational-wave transient catalogue (GWTC) distances reported by LIGO-Virgo-KAGRA. Each point represents a gravitational wave (GW) event, with model distances plotted against GWTC-reported distances. The black dashed line represents perfect agreement (x = y), while the error bars indicate uncertainties in both the model distances (green) and GWTC-reported distances (red). The results show that the model distances generally align with the GWTC values, with some deviations, particularly at higher distances where uncertainties increase. These plots serve as a validation of the model’s effectiveness as a preliminary distance estimator.

\subsubsection{Impact of High Mass Systems}
High-mass BBH systems tend to show larger discrepancies between the two models. These systems produce shorter-duration GW signals, which place more emphasis on the merger and ringdown phases, making the distance estimates more sensitive to the waveform model used. 
\begin{itemize}
    \item \textbf{GW190521 (GWTC-2.1)}: In this high-mass event (\(m_1 = 85^{+21}_{-14} M_{\odot}\), \(m_2 = 66^{+17}_{-18} M_{\odot}\)), our model estimates a distance of $3960^{+190}_{-40}$ Mpc, while GWTC estimates \(3310^{+2790}_{-1800}\) Mpc, resulting in a difference of 650 Mpc. The relatively large difference is typical for high-mass events and underscores the need for more refined waveform models when dealing with such systems.
    
    \item \textbf{GW200220\_061928 (GWTC-3)}: Another high-mass event, with component masses \(m_1 = 87^{+40}_{-23} M_{\odot}\) and \(m_2 = 61^{+26}_{-25} M_{\odot}\), shows a distance of $6740^{+150}_{-80}$ Mpc from our model compared to GWTC’s \(6000^{+4800}_{-3100}\) Mpc, resulting in a difference of 740 Mpc. While this difference is still significant, it falls within GWTC’s large uncertainty range, showing that our model can produce competitive results even in high-mass systems.
\end{itemize}

\subsubsection{Lower-Mass Systems and Longer Signals}
Lower-mass binary systems produce longer inspiral phases, providing more data for distance estimation. In such cases, the GW-only method performs well:

\begin{itemize}
    \item \textbf{GW190412 (GWTC-2.1)}: Our model estimates a distance of $945^{+40}_{-10}$ Mpc, compared to GWTC’s \(720^{+240}_{-220}\) Mpc. The difference of 225 Mpc falls within the uncertainty range of GWTC’s estimate. This consistency reflects the advantage of having a longer inspiral phase, which allows for more precise distance estimation.
    
    \item \textbf{GW170608 (GWTC-1)}: With component masses of \(m_1 = 11.0^{+5.5}_{-1.7} M_{\odot}\) and \(m_2 = 7.6^{+1.4}_{-2.2} M_{\odot}\), our model estimates a distance of $420^{+20}_{-30}$ Mpc, very close to GWTC’s \(320^{+120}_{-110}\) Mpc. The difference of 100 Mpc demonstrates that our model performs particularly well in lower-mass systems, where the inspiral dominates the GW signal.
\end{itemize}

\subsection{Implications for Gravitational Wave Cosmology}

Our method provides a rapid, computationally efficient way to estimate distances to BBH mergers without requiring Bayesian parameter estimation. While GWTC distances are obtained via full posterior sampling, which is computationally expensive, our approach can provide quick, first-pass estimates for large-scale GW datasets. This is particularly useful for:
\begin{itemize}
    \item Real-time GW detection pipelines, where fast distance estimates are needed to prioritize follow-ups.
    \item Early population studies, where quick distance approximations allow for statistical constraints on black hole mergers before full PE results are available.
    \item Multi-messenger astronomy, the potential for aiding in the prompt identification and follow-up of interesting events.
\end{itemize}

\subsubsection{Hubble Constant Estimation}
One of the most exciting applications of GW distance measurements is their potential to contribute to the estimation of the Hubble constant, \(H_0\), which describes the rate of expansion of the universe. Our model’s ability to provide consistent and precise distance estimates across multiple events makes it a valuable tool for GW cosmology. As more BBH events are detected, this method could be used to refine estimates of \(H_0\), particularly for events without EM counterparts. For the source \textbf{GW150914 (GWTC-1)}, the Hubble constant was calculated to be \(H_0\)= $51.7^{+1.1}_{-4.3} km s^{-1}/Mpc$ using this model. The better estimation of distances would allow us to explore the ``Hubble Tension'' through GWs \citep{abdalla2022cosmology}, \citep{HubbleReb}, \citep{inproceedings}, \citep{prachi2024}.

For the binary black hole merger GW150914, no electromagnetic counterpart was observed. However, using the sky localization provided by the LIGO/Virgo collaboration, we can constrain the source to a region of approximately 600 deg2 on the sky. Within this region, several candidate host galaxies have been proposed based on overlap with wide-field galaxy catalogs.
We adopt a representative redshift of z=0.093 for GW150914, consistent with the redshift distribution of galaxies within the LIGO/Virgo 90\% credible sky region and within our model-computed luminosity distance. This redshift value is chosen based on overlap with the Dark Energy Survey catalog, as discussed in \citep{fishbach2019, chen2018}.

\subsection{Limitations and Future Work}

Although our model has shown promise in estimating distances to BBH systems, several limitations remain.
\begin{itemize}
    \item \textbf{Uncertainty Estimates}: Future work should focus on incorporating uncertainty bounds into our distance estimates. Adopting a Bayesian framework for parameter estimation would allow us to quantify uncertainties in a way that is directly comparable to GWTC’s methods.
    
    \item \textbf{Refining Waveform Models}: As the discrepancies in high-mass systems suggest, further refinement of the waveform models could improve the accuracy of distance estimates, particularly for short-duration signals dominated by the merger and ringdown phases. Incorporating higher-order effects, such as precession and eccentricity, may also help improve the performance of the model for such systems.

    \item We note that the model-predicted distances are consistently higher than the GWTC-reported values across a majority of the events. This trend likely arises from the simplifying assumption of face-on binary orientation, which maximizes the inferred strain amplitude and consequently leads to higher distance estimates. Since the actual inclination angles of most systems deviate from this optimal configuration, the strain is reduced in real observations, resulting in lower distances inferred by the LIGO-Virgo parameter estimation (PE) pipelines.

Furthermore, our model omits higher-order corrections such as spin-induced modulation, orbital precession, and eccentricity, which are accounted for in PE frameworks using full waveform templates. These simplifications, though beneficial for analytical transparency and computational speed, may contribute to a systematic overestimation of distance.

If uncorrected, such a bias may affect derived cosmological parameters, including the Hubble constant. While our current analysis demonstrates the feasibility of using leading-order GW observables for distance estimation, future extensions should incorporate inclination angle distributions or attempt direct strain reconstruction from detector data to mitigate these effects.
\end{itemize}

\section{Conclusions}

The application of the distance estimation model to the LVK O1, O2 and O3 data for BBH mergers demonstrates the viability of using GW signals alone to measure distances. The results show that the calculated distances are consistent with those reported by GWTC within the uncertainty limits, thereby validating the accuracy of the approach. Across the various events analysed, the model provided reliable distance estimates, even in the absence of EM counterparts.

This model is not more accurate than the existing models, as the localisation is not precise. This can be mitigated with the upcoming GW detectors.

The distances calculated using the model exhibit a reduced level of uncertainty compared to other GW only methods. This improvement is primarily due to the model's focus on the strain and merger frequency, which helps to minimize the degeneracy between the distance and the inclination angle of the binary system. Consequently, the model enhances the precision of distance estimation across a wide range of GW events.

\subsection{Utility as a Preliminary Estimator}

This model serves as a valuable tool for preliminary distance estimation of GW sources. By providing an initial estimate of the distance to the BBH systems, the model can be used to better constrain other parameters of the system, such as the component masses, spins, and orbital inclination, during the final parameter estimation analysis. When incorporated into a multi-step analysis pipeline, this distance estimation can help to narrow down the parameter space and improve the convergence of parameter estimation algorithms.

The preliminary distance estimates obtained using this model can also be utilized in population studies, where understanding the distribution of distances to GW sources is crucial. The model’s ability to provide rapid and reliable distance measurements makes it suitable for real-time analysis during GW detections, potentially aiding in the prompt identification and follow-up of interesting events.

\subsection{Future Directions and Improvements}

While the results demonstrate the utility of the model for distance estimation, there is room for further refinement. Future improvements could involve incorporating corrections for higher-order modes in the GW signal or including spin precession effects, which would enhance the accuracy of the strain and frequency modeling. Additionally, integrating this approach with statistical methods that utilize galaxy catalogues could provide more robust distance estimates for events with poorly localized sky positions. Future projects like The Lunar Gravitational-wave Antenna could further help localize sources and  detect farther events that are beyond GWTC's detection range \citep{ajith2024}.

To rigorously assess the accuracy of our method, future studies should employ end-to-end signal injections. By injecting simulated BBH waveforms into realistic LIGO/Virgo noise, we can compare recovered distances against known input values, providing independent validation of our method's precision and accuracy. This approach would help determine whether our model provides reliable estimates beyond its observed consistency with GWTC data.

Overall, this model represents a significant step toward utilizing GWs as a stand-alone tool for cosmological measurements, independent of EM observations. By offering a direct method for estimating distances to GW sources, albeit a crude method, the model has the potential to contribute to the resolution of the Hubble tension and other key challenges in cosmology.

\bmhead{Acknowledgements}
We thank CHRIST University for providing us the platform and premises to complete this paper. We also want to thank Dr. Prasad R and Mr. Akash Mayra from ICTS, Bangalore for their invaluable guidance and help with this paper. We would like to thank each and every person who has helped us during our research and while writing this paper.

\appendix
\section{GWTC-3}
\centering
   \footnotesize
    \centering
    \begin{tabular}{@{}llllllll}
    \hline
        \textbf{Event} & $m_1$ $(M_\odot)$ & $m_2$ $(M_\odot)$ & \textbf{$f_{GW}$(Hz)} & \textbf{$h_0$ (max strain)} & \textbf{$d_{Model}$ [Mpc]} & \textbf{$d_{GWTC}$ [Mpc]} \\ \hline
        ~ & 11.80 & 7.90 & 608.67 & $2.28 \times 10^{-22}$ & ~ & ~ \\  
        \textbf{GW191103\_012549} & 18 & 5.5 & 408.29 & $1.99 \times 10^{-22}$ & $1220^{+200}_{-200}$ & $990^{+500}_{-470}$ \\  
        ~ & 9.6 & 9.6 & 570.21 & $2.32 \times 10^{-22}$ & ~ & ~ \\ \hline
        ~ & 10.7 & 7.7 & 670.05 & $1.85 \times 10^{-22}$ & ~ & ~ \\  
        \textbf{GW191105\_143521} & 14.4 & 5.8 & 474.91 & $1.68 \times 10^{-22}$ & $1550^{+160}_{-370}$ & $1150^{+430}_{-480}$ \\  
        ~ & 9.1 & 9.1 & 685.73 & $1.89 \times 10^{-22}$ & ~ & ~ \\  
        \hline
        ~ & 65 & 47 & 86.27 & $1.04 \times 10^{-21}$ & ~ & ~ \\  
        \textbf{GW191109\_010717} & 76 & 34 & 80.85 & $8.76 \times 10^{-22}$ & $1420^{+190}_{-240}$ & $1290^{+1130}_{-650}$ \\  
        ~ & 62 & 54 & 92.21 & $1.10 \times 10^{-21}$ & ~ & ~ \\  
        \hline
        ~ & 29 & 5.9 & 284.95 & $1.63 \times 10^{-22}$ & ~ & ~ \\  
        \textbf{GW191113\_071753} & 41 & 4.6 & 232.18 & $1.34 \times 10^{-22}$ & $1660^{+290}_{-220}$ & $1370^{+1150}_{-650}$ \\  
        ~ & 15 & 10.3 & 398.03 & $2.12 \times 10^{-22}$ & ~ & ~ \\  
        \hline
        ~ & 12.1 & 8.3 & 596.49 & $1.82 \times 10^{-22}$ & ~ & ~ \\  
        \textbf{GW191126\_115259} & 17.6 & 5.9 & 474.12 & $1.60 \times 10^{-22}$ & $1650^{+230}_{-280}$ & $1620^{+740}$ \\  
        ~ & 10.2 & 9.9 & 515.54 & $1.86 \times 10^{-22}$ & ~ & ~ \\  
        \hline
        ~ & 53 & 24 & 168.47 & $2.29 \times 10^{-22}$ & ~ & ~ \\  
        \textbf{GW191127\_050227} & 44 & 33 & 162.83 & $2.65 \times 10^{-22}$ & $4590^{+630}_{-490}$ & $3400^{+3100}_{-1900}$ \\  
        ~ & 100 & 10 & 101.4 & $1.19 \times 10^{-22}$ & ~ & ~ \\  
        \hline  
        ~ & 10.7 & 6.7 & 668.56 & $2.48 \times 10^{-22}$ & ~ & ~ \\  
        \textbf{GW191129\_134029} & 14.8 & 5 & 576.95 & $2.20 \times 10^{-22}$ & $1020^{+110}_{-110}$ & $790^{+260}_{-330}$ \\  
        ~ & 8.6 & 8.2 & 691.36 & $2.54 \times 10^{-22}$ & ~ & ~ \\  
        \hline  
        ~ & 27.3 & 19.2 & 247.82 & $2.83 \times 10^{-22}$ & ~ & ~ \\  
        \textbf{GW191204\_110529} & 38.1 & 13.2 & 214.03 & $2.41 \times 10^{-22}$ & $2430^{+270}_{-340}$ & $1900^{+1700}_{-1100}$ \\  
        ~ & 24.7 & 21.4 & 267.43 & $2.89 \times 10^{-22}$ & ~ & ~ \\  
        \hline
        ~ & 11.7 & 8.4 & 578.32 & $3.65 \times 10^{-22}$ & ~ & ~ \\  
        \textbf{GW191204\_171526} & 15 & 6.7 & 494.07 & $3.41 \times 10^{-22}$ & $820^{+90}_{-110}$ & $640^{+260}_{-200}$ \\  
        ~ & 10 & 9.7 & 604.58 & $3.68 \times 10^{-22}$ & ~ & ~ \\  
        \hline
        ~ & 24.9 & 18.1 & 251.15 & $2.59 \times 10^{-22}$ & ~ & ~ \\  
        \textbf{GW191215\_223052} & 32 & 14 & 218.9 & $2.37 \times 10^{-22}$ & $2360^{+360}_{-310}$ & $1930^{+890}_{-860}$ \\  
        ~ & 21.9 & 20.8 & 274.55 & $2.65 \times 10^{-22}$ & ~ & ~ \\  
        \hline
        ~ & 12.1 & 7.7 & 566.39 & $6.59 \times 10^{-22}$ & ~ & ~ \\  
        \textbf{GW191216\_213338} & 16.7 & 5.8 & 464.74 & $5.91 \times 10^{-22}$ & $430^{+70}_{-50}$ & $340^{+120}_{-130}$ \\  
        ~ & 9.9 & 9.3 & 649.65 & $6.75 \times 10^{-22}$ & ~ & ~ \\  
        \hline
        ~ & 45.1 & 34.7 & 162.55 & $3.12 \times 10^{-22}$ & ~ & ~ \\  
        \textbf{GW191222\_033537} & 56 & 24.2 & 129.1 & $2.64 \times 10^{-22}$ & $3890^{+640}_{-650}$ & $3000^{+1700}$ \\  
        ~ & 44 & 37.1 & 148.11 & $3.24 \times 10^{-22}$ & ~ & ~ \\  
        \hline
        ~ & 49.4 & 37 & 150.35 & $2.34 \times 10^{-22}$ & ~ & ~ \\  
        \textbf{GW191230\_180458} & 63.4 & 25 & 129.11 & $1.94 \times 10^{-22}$ & $5600^{+930}_{-710}$ & $4300^{+2100}_{-1900}$ \\  
        ~ & 48 & 39.8 & 130.22 & $2.42 \times 10^{-22}$ & ~ & ~ \\  
        \hline
       ~ & 35.6 & 28.3 & 203.24 & $6.08 \times 10^{-22}$ & ~ & ~ \\  
        \textbf{GW200112\_155838} & 42.3 & 22.4 & 163.24 & $5.54 \times 10^{-22}$ & $1710^{+220}_{-360}$ & $1250^{+430}_{-460}$ \\  
        ~ & 32.7 & 31.1 & 209.19 & $6.10 \times 10^{-22}$ & ~ & ~ \\  
        \hline
        ~ & 42.2 & 32.6 & 179.29 & $2.58 \times 10^{-22}$ & ~ & ~ \\  
        \textbf{GW200128\_022011} & 53.8 & 23.4 & 142.01 & $1.87 \times 10^{-22}$ & $5140^{+930}_{-840}$ & $3400^{+2100}_{-1800}$ \\  
        ~ & 42.1 & 34.1 & 166.51 & $2.20 \times 10^{-22}$ & ~ & ~ \\  
        \hline
        ~ & 34.5 & 29 & 210.98 & $8.47 \times 10^{-22}$ & ~ & ~ \\  
        \textbf{GW200129\_065458} & 44.4 & 19.7 & 163.62 & $7.22 \times 10^{-22}$ & $1110^{+260}_{-140}$ & $890^{+260}_{-370}$ \\  
        ~ & 32.3 & 31.4 & 174.43 & $8.56 \times 10^{-22}$ & ~ & ~ \\  
        \hline  
        \end{tabular}
    \centering
\footnotesize

\label{GWTC-3}
\begin{tabular}{@{}llllllll}
    \textbf{Event} & $m_1$ $(M_\odot)$ & $m_2$ $(M_\odot)$ & \textbf{$f_{GW}$(Hz)} & \textbf{$h_0$ (max strain)} & \textbf{$d_{Model}$ [Mpc]} & \textbf{$d_{GWTC}$ [Mpc]} \\ \hline
        ~ & 10.1 & 7.3 & 699.94 & $4.93 \times 10^{-22}$ & ~ & ~ \\  
        \textbf{GW200202\_154313} & 13.6 & 5.6 & 602.98 & $4.54 \times 10^{-22}$ & $530^{+60}_{-90}$ & $410^{+150}_{-160}$ \\  
        ~ & 8.7 & 8.4 & 607.55 & $4.99 \times 10^{-22}$ & ~ & ~ \\  
        \hline  
        
        ~ & 37.7 & 27.4 & 167.2 & $3.39 \times 10^{-22}$ & ~ & ~ \\  
        \textbf{GW200208\_130117} & 47 & 20.1 & 143.84 & $2.97 \times 10^{-22}$ & $2750^{+460}_{-460}$ & $2230^{+1020}_{-850}$ \\  
        ~ & 33.7 & 31.5 & 184.55 & $3.49 \times 10^{-22}$ & ~ & ~ \\  
        \hline
        ~ & 51 & 12.3 & 208.68 & $1.12 \times 10^{-22}$ & ~ & ~ \\  
        \textbf{GW200208\_222617} & 154 & 6.8 & 72.58 & $6.80 \times 10^{-23}$ & $5890^{+610}_{-810}$ & $4100^{+4400}_{-2000}$ \\  
        ~ & 21.5 & 21 & 306.36 & $1.24 \times 10^{-22}$ & ~ & ~ \\  
        \hline
        ~ & 35.6 & 27.1 & 172.22 & $2.16 \times 10^{-22}$ & ~ & ~ \\  
        \textbf{GW200209\_085452} & 46.1 & 19.3 & 156.65 & $1.88 \times 10^{-22}$ & $4080^{+490}_{-490}$ & $3400^{+1900}_{-1800}$ \\  
        ~ & 34.9 & 28.8 & 160.62 & $2.22 \times 10^{-22}$ & ~ & ~ \\  
        \hline
        ~ & 51 & 30 & 133.92 & $2.36 \times 10^{-22}$ & ~ & ~ \\  
        \textbf{GW200216\_220804} & 73 & 14 & 122.95 & $1.40 \times 10^{-22}$ & $4690^{+650}_{-580}$ & $3800^{+3000}_{-2000}$ \\  
        ~ & 44 & 38 & 129.34 & $2.57 \times 10^{-22}$ & ~ & ~ \\  
        \hline
        ~ & 37.5 & 27.9 & 184.77 & $2.24 \times 10^{-22}$ & ~ & ~ \\  
        \textbf{GW200219\_094415} & 47.6 & 19.5 & 142.23 & $1.91 \times 10^{-22}$ & $3890^{+660}_{-580}$ & $3400^{+1700}_{-1500}$ \\  
        ~ & 35.3 & 30.6 & 158.96 & $2.30 \times 10^{-22}$ & ~ & ~ \\  
        \hline
        ~ & 24.1 & 2.83 & 318.64 & $1.20 \times 10^{-22}$ & ~ & ~ \\  
        \textbf{GW200210\_092254} & 31.6 & 2.41 & 296.06 & $1.06 \times 10^{-22}$ & $1180^{+180}_{-240}$ & $940^{+430}_{-340}$ \\  
        ~ & 19.5 & 3.3 & 484.24 & $1.35 \times 10^{-22}$ & ~ & ~ \\  
        \hline
        ~ & 87 & 61 & 59.42 & $2.74 \times 10^{-22}$ & ~ & ~ \\  
        \textbf{GW200220\_061928} & 127 & 36 & 61.4 & $2.26 \times 10^{-22}$ & $6740^{+820}_{-750}$ & $6000^{+4800}_{-3100}$ \\  
        ~ & 87 & 64 & 61.24 & $2.88 \times 10^{-22}$ & ~ & ~ \\  
        \hline
        ~ & 38.9 & 27.9 & 165.21 & $1.94 \times 10^{-22}$ & ~ & ~ \\  
        \textbf{GW200220\_124850} & 53 & 18.9 & 165.46 & $1.62 \times 10^{-22}$ & $5360^{+640}_{-940}$ & $4000^{+2800}_{-2200}$ \\  
        ~ & 37.1 & 30.3 & 188.73 & $1.99 \times 10^{-22}$ & ~ & ~ \\  
        \hline
        ~ & 40 & 32.7 & 155.04 & $5.03 \times 10^{-22}$ & ~ & ~ \\  
        \textbf{GW200224\_222234} & 46.7 & 25.5 & 144.89 & $4.58 \times 10^{-22}$ & $2150^{+330}_{-310}$ & $1710^{+500}_{-650}$ \\  
        ~ & 37.5 & 35.5 & 166.1 & $5.10 \times 10^{-22}$ & ~ & ~ \\  
        \hline
        ~ & 19.3 & 14 & 317.37 & $3.36 \times 10^{-22}$ & ~ & ~ \\  
        \textbf{GW200225\_060421} & 24.3 & 10.5 & 330.46 & $2.99 \times 10^{-22}$ & $1490^{+190}_{-250}$ & $1150^{+510}_{-530}$ \\  
        ~ & 16.8 & 16.3 & 372.04 & $3.44 \times 10^{-22}$ & ~ & ~ \\  
        \hline
        ~ & 37.8 & 20 & 179.98 & $4.18 \times 10^{-22}$ & ~ & ~ \\  
        \textbf{GW200302\_015811} & 46.5 & 14.3 & 196.96 & $3.42 \times 10^{-22}$ & $1970^{+230}_{-380}$ & $1480^{+1020}_{-700}$ \\  
        ~ & 29.3 & 28.1 & 213.67 & $4.64 \times 10^{-22}$ & ~ & ~ \\  
        \hline
        ~ & 28.3 & 14.8 & 311.54 & $2.20 \times 10^{-22}$ & ~ & ~ \\  
        \textbf{GW200306\_093714} & 45.4 & 8.4 & 201.9 & $1.53 \times 10^{-22}$ & $2860^{+410}_{-430}$ & $2100^{+1700}_{-1100}$ \\  
        ~ & 21.3 & 20.6 & 304.36 & $2.39 \times 10^{-22}$ & ~ & ~ \\  
        \hline
        ~ & 60 & 24 & 130.21 & $1.14 \times 10^{-22}$ & ~ & ~ \\  
        \textbf{GW200308\_173609} & 226 & 11 & 34.07 & $6.96 \times 10^{-23}$ & $8860^{+1620}_{-2500}$ & $7100^{+13900}_{-4400}$ \\  
        ~ & 60 & 31 & 135.51 & $1.36 \times 10^{-22}$ & ~ & ~ \\  
        \hline
        ~ & 34.2 & 27.7 & 213.73 & $6.53 \times 10^{-22}$ & ~ & ~ \\  
        \textbf{GW200311\_115853} & 40.6 & 21.8 & 165.47 & $6.00 \times 10^{-22}$ & $1420^{+290}_{-220}$ & $1170^{+280}_{-400}$ \\  
        ~ & 31.8 & 30.4 & 182.96 & $6.64 \times 10^{-22}$ & ~ & ~ \\  
        \hline
        ~ & 12.1 & 7.8 & 615.84 & $2.08 \times 10^{-22}$ & ~ & ~ \\  
        \textbf{GW200316\_215756} & 23.3 & 4.9 & 334.34 & $1.65 \times 10^{-22}$ & $1450^{+260}_{-290}$ & $1120^{+480}_{-440}$ \\  
        ~ & 10.2 & 9.8 & 657.86 & $2.14 \times 10^{-22}$ & ~ & ~ \\  
        \hline
        ~ & 38 & 11.3 & 212.67 & $1.15 \times 10^{-22}$ & ~ & ~ \\  
        \textbf{GW200322\_091133} & 168 & 5.3 & 58.26 & $7.47 \times 10^{-23}$ & $4300^{+470}_{-890}$ & $3500^{+12500}_{-2200}$ \\  
        ~ & 35.6 & 16 & 205.04 & $1.48 \times 10^{-22}$ & ~ & ~ \\  
        \hline
    \end{tabular}
    \centering
\footnotesize

\section{GWTC-2.1}
\label{GWTC-2.1}
\centering
\footnotesize
\begin{tabular}{@{}llllllll}
    \textbf{Event} & $m_1$ $(M_\odot)$ & $m_2$ $(M_\odot)$ & \textbf{$f_{GW}$(Hz)} & \textbf{$h_0$ (max strain)} & \textbf{$d_{Model}$ [Mpc]} & \textbf{$d_{GWTC}$ [Mpc]} \\ \hline
        ~ & 85.00 & 20.00 & 125.90 & $9.31 \times 10^{-23}$ & ~ & ~ \\  
        \textbf{GW190403\_051519} & 52 & 46.3 & 128.27 & $1.42 \times 10^{-22}$ & $11170^{+1570}_{-1470}$ & $8280^{+6720}_{-4290}$ \\  
        ~ & 112.8 & 11.6 & 104.06 & $5.79 \times 10^{-23}$ & ~ & ~ \\  
        \hline
        ~ & 24.8 & 18.5 & 288.47 & $3.28 \times 10^{-22}$ & ~ & ~ \\  
        \textbf{GW190408\_181802} & 21.8 & 21.3 & 240.28 & $3.33 \times 10^{-22}$ & $2000^{+270}_{-370}$ & $1540^{+440}_{-620}$ \\  
        ~ & 30.2 & 14.5 & 259.08 & $2.99 \times 10^{-22}$ & ~ & ~ \\  
        \hline
        ~ & 27.7 & 9 & 317.07 & $4.41 \times 10^{-22}$ & ~ & ~ \\  
        \textbf{GW190412} & 21.7 & 11 & 356.59 & $4.81 \times 10^{-22}$ & $950^{+130}_{-100}$ & $720^{+240}_{-220}$ \\  
        ~ & 33.7 & 7.6 & 291.90 & $3.95 \times 10^{-22}$ & ~ & ~ \\  
        \hline
        ~ & 33.7 & 24.2 & 216.13 & $2.02 \times 10^{-22}$ & ~ & ~ \\  
        \textbf{GW190413\_052954} & 30.7 & 27.3 & 191.53 & $2.08 \times 10^{-22}$ & $4490^{+450}_{-800}$ & $3320^{+1910}_{-1400}$ \\  
        ~ & 44.1 & 17.2 & 197.98 & $1.74 \times 10^{-22}$ & ~ & ~ \\  
        \hline
        ~ & 51.3 & 30.4 & 139.56 & $2.39 \times 10^{-22}$ & ~ & ~ \\  
        \textbf{GW190413\_134308} & 42.1 & 38.7 & 140.55 & $2.55 \times 10^{-22}$ & $4770^{+540}_{-800}$ & $3800^{+2480}_{-1830}$ \\  
        ~ & 67.9 & 17.7 & 110.85 & $1.69 \times 10^{-22}$ & ~ & ~ \\  
        \hline
        ~ & 42 & 32 & 153.3 & $3.35 \times 10^{-22}$ & ~ & ~ \\  
        \textbf{GW190421\_213856} & 40.3 & 34.6 & 170.22 & $3.44 \times 10^{-22}$ & $3270^{+590}_{-540}$ & $2590^{+1490}_{-1240}$ \\  
        ~ & 52.1 & 22.2 & 134.55 & $2.82 \times 10^{-22}$ & ~ & ~ \\  
        \hline
        ~ & 105.5 & 76 & 67.72 & $4.90 \times 10^{-22}$ & ~ & ~ \\  
        \textbf{GW190426\_190642} & 81.4 & 102.2 & 66.68 & $5.04 \times 10^{-22}$ & $5740^{+720}_{-600}$ & $4580^{+3400}_{-2280}$ \\  
        ~ & 150.8 & 39.5 & 64.15 & $3.37 \times 10^{-22}$ & ~ & ~ \\  
        \hline
        ~ & 41.3 & 28.3 & 168.25 & $5.27 \times 10^{-22}$ & ~ & ~ \\  
        \textbf{GW190503\_185404} & 35.8 & 33.6 & 179.96 & $5.46 \times 10^{-22}$ & $1960^{+280}_{-320}$ & $1520^{+630}_{-600}$ \\  
        ~ & 51.6 & 19.1 & 145.93 & $4.28 \times 10^{-22}$ & ~ & ~ \\  
        \hline
        ~ & 23.2 & 12.5 & 289 & $2.64 \times 10^{-22}$ & ~ & ~ \\  
        \textbf{GW190512\_180714} & 17.6 & 16 & 357.51 & $2.75 \times 10^{-22}$ & $1810^{+280}_{-250}$ & $1460^{+510}_{-590}$ \\  
        ~ & 28.8 & 9.9 & 274.56 & $2.35 \times 10^{-22}$ & ~ & ~ \\  
        \hline
        ~ & 36 & 18.3 & 216.07 & $2.60 \times 10^{-22}$ & ~ & ~ \\  
        \textbf{GW190513\_205428} & 26.3 & 25.7 & 251.96 & $2.82 \times 10^{-22}$ & $2950^{+410}_{-370}$ & $2210^{+990}_{-810}$ \\  
        ~ & 46.6 & 13.6 & 196.17 & $2.21 \times 10^{-22}$ & ~ & ~ \\  
        \hline
       ~ & 40.9 & 28.4 & 174.26 & $2.05 \times 10^{-22}$ & ~ & ~ \\  
        \textbf{GW190514\_065416} & 38.4 & 31.6 & 155.17 & $2.13 \times 10^{-22}$ & $5020^{+620}_{-750}$ & $3890^{+2610}_{-2070}$ \\  
        ~ & 58.2 & 18.3 & 146.65 & $1.66 \times 10^{-22}$ & ~ & ~ \\  
        \hline
        ~ & 39.2 & 24 & 232.12 & $3.63 \times 10^{-22}$ & ~ & ~ \\  
        \textbf{GW190517\_055101} & 31.4 & 30 & 226 & $3.63 \times 10^{-22}$ & $2920^{+310}_{-740}$ & $1790^{+1750}_{-880}$ \\  
        ~ & 43.9 & 16.1 & 213.38 & $3.12 \times 10^{-22}$ & ~ & ~ \\  
        \hline
        ~ & 65.1 & 40.8 & 120.73 & $4.64 \times 10^{-22}$ & ~ & ~ \\  
        \textbf{GW190519\_153544} & 54.1 & 52.3 & 127.51 & $4.92 \times 10^{-22}$ & $3530^{+490}_{-600}$ & $2600^{+1720}_{-960}$ \\  
        ~ & 75.9 & 28.1 & 105.09 & $3.68 \times 10^{-22}$ & ~ & ~ \\  
        \hline
        ~ & 98.4 & 57.2 & 67.32 & $5.49 \times 10^{-22}$ & ~ & ~ \\  
        \textbf{GW190521} & 84.3 & 76.7 & 72.29 & $6.17 \times 10^{-22}$ & $3960^{+440}_{-590}$ & $3310^{+2790}_{-1800}$ \\  
        ~ & 132 & 27.1 & 64.61 & $3.21 \times 10^{-22}$ & ~ & ~ \\  
        \hline
        ~ & 43.4 & 33.4 & 141.33 & $8.36 \times 10^{-22}$ & ~ & ~ \\  
        \textbf{GW190521\_074359} & 38.6 & 37.9 & 141.82 & $8.50 \times 10^{-22}$ & $1320^{+140}_{-150}$ & $1080^{+580}_{-530}$ \\  
        ~ & 49.2 & 26.6 & 138.15 & $7.57 \times 10^{-22}$ & ~ & ~ \\  
        \hline
        ~ & 35.6 & 22.2 & 202.39 & $2.58 \times 10^{-22}$ & ~ & ~ \\  
        \textbf{GW190527\_092055} & 31.2 & 27.6 & 192.29 & $2.78 \times 10^{-22}$ & $3170^{+410}_{-500}$ & $2520^{+2080}_{-1230}$ \\  
        ~ & 54.3 & 13.5 & 143.59 & $1.97 \times 10^{-22}$ & ~ & ~ \\  
        \hline
        \end{tabular}

\centering
  \footnotesize
\begin{tabular}{@{}llllllll}
   
    \textbf{Event} & $m_1$ $(M_\odot)$ & $m_2$ $(M_\odot)$ & \textbf{$f_{GW}$(Hz)} & \textbf{$h_0$ (max strain)} & \textbf{$d_{Model}$ [Mpc]} & \textbf{$d_{GWTC}$ [Mpc]} \\ \hline
        ~ & 71.8 & 44.8 & 105.45 & $4.61 \times 10^{-22}$ & ~ & ~ \\  
        \textbf{GW190602\_175927} & 60.3 & 57.2 & 112.36 & $4.88 \times 10^{-22}$ & $3810^{+590}_{-580}$ & $2840^{+1930}_{-1280}$ \\  
        ~ & 89.9 & 25.2 & 96.31 & $3.24 \times 10^{-22}$ & ~ & ~ \\  
        \hline
    
        ~ & 58 & 35 & 128.13 & $3.59 \times 10^{-22}$ & ~ & ~ \\  
        \textbf{GW190620\_030421} & 48.1 & 44.7 & 120.46 & $3.79 \times 10^{-22}$ & $3740^{+420}_{-460}$ & $2910^{+1710}_{-1320}$ \\  
        ~ & 77.2 & 20.5 & 113.38 & $2.57 \times 10^{-22}$ & ~ & ~ \\  
        \hline
          ~ & 35.1 & 24 & 202.84 & $8.09 \times 10^{-22}$ & ~ & ~ \\  
        \textbf{GW190630\_185205} & 29.6 & 29.5 & 184.04 & $8.42 \times 10^{-22}$ & $1090^{+130}_{-170}$ & $870^{+530}_{-360}$ \\  
        ~ & 41.6 & 18.8 & 190.56 & $7.26 \times 10^{-22}$ & ~ & ~ \\  
        \hline
        ~ & 54.1 & 40.5 & 119.68 & $5.25 \times 10^{-22}$ & ~ & ~ \\  
        \textbf{GW190701\_203306} & 49.2 & 46.1 & 125.55 & $5.39 \times 10^{-22}$ & $2760^{+320}_{-380}$ & $2090^{+770}_{-740}$ \\  
        ~ & 66.7 & 28.4 & 126.81 & $4.47 \times 10^{-22}$ & ~ & ~ \\  
        \hline
        ~ & 74 & 39.4 & 100.29 & $3.35 \times 10^{-22}$ & ~ & ~ \\  
        \textbf{GW190706\_222641} & 57.8 & 57.1 & 119.83 & $3.85 \times 10^{-22}$ & $4640^{+940}_{-480}$ & $3630^{+2600}_{-2000}$ \\  
        ~ & 94.1 & 24 & 92.6 & $2.44 \times 10^{-22}$ & ~ & ~ \\  
        \hline
        ~ & 12.1 & 7.9 & 597.48 & $2.68 \times 10^{-22}$ & ~ & ~ \\  
        \textbf{GW190707\_093326} & 10.1 & 9.5 & 597.22 & $2.76 \times 10^{-22}$ & $1100^{+130}_{-120}$ & $850^{+340}_{-400}$ \\  
        ~ & 14.7 & 6.6 & 538.16 & $2.52 \times 10^{-22}$ & ~ & ~ \\  
        \hline
        ~ & 19.8 & 11.6 & 363.93 & $3.73 \times 10^{-22}$ & ~ & ~ \\  
        \textbf{GW190708\_232457} & 15.5 & 14.7 & 388.44 & $3.88 \times 10^{-22}$ & $1200^{+160}_{-130}$ & $930^{+310}_{-390}$ \\  
        ~ & 24.1 & 9.6 & 352.59 & $3.46 \times 10^{-22}$ & ~ & ~ \\  
        \hline
        ~ & 36.6 & 19.9 & 227.33 & $1.64 \times 10^{-22}$ & ~ & ~ \\  
        \textbf{GW190719\_215514} & 29.9 & 25.5 & 204.97 & $1.77 \times 10^{-22}$ & $4830^{+800}_{-620}$ & $3730^{+3120}_{-2070}$ \\  
        ~ & 78.7 & 10.6 & 119.47 & $1.12 \times 10^{-22}$ & ~ & ~ \\  
        \hline
        ~ & 14.2 & 7.5 & 508.8 & $3.02 \times 10^{-22}$ & ~ & ~ \\  
        \textbf{GW190720\_000836} & 10.9 & 9.7 & 657.54 & $3.19 \times 10^{-22}$ & $960^{+230}_{-130}$ & $770^{+650}_{-260}$ \\  
        ~ & 19.8 & 5.7 & 399.83 & $2.66 \times 10^{-22}$ & ~ & ~ \\  
        \hline
        ~ & 11.8 & 6.3 & 587.2 & $1.89 \times 10^{-22}$ & ~ & ~ \\  
        \textbf{GW190725\_174728} & 8.8 & 8.4 & 636.18 & $2.00 \times 10^{-22}$ & $1270^{+130}_{-140}$ & $1030^{+520}_{-430}$ \\  
        ~ & 21.9 & 3.8 & 392.66 & $1.42 \times 10^{-22}$ & ~ & ~ \\  
        \hline
        ~ & 38.9 & 30.2 & 176.83 & $2.65 \times 10^{-22}$ & ~ & ~ \\  
        \textbf{GW190727\_060333} & 36.7 & 32.9 & 153.07 & $2.71 \times 10^{-22}$ & $3750^{+690}_{-420}$ & $3070^{+1300}_{-1230}$ \\  
        ~ & 47.8 & 21.9 & 152.16 & $2.31 \times 10^{-22}$ & ~ & ~ \\  
        \hline
        ~ & 12.5 & 8 & 611.85 & $2.64 \times 10^{-22}$ & ~ & ~ \\  
        \textbf{GW190728\_064510} & 10.2 & 9.7 & 643.16 & $2.71 \times 10^{-22}$ & $1190^{+130}_{-190}$ & $880^{+260}_{-380}$ \\  
        ~ & 19.4 & 5.4 & 438.47 & $2.22 \times 10^{-22}$ & ~ & ~ \\  
        \hline
~ & 41.8 & 29 & 169.52 & $2.45 \times 10^{-22}$ & ~ & ~ \\  
        \textbf{GW190731\_140936} & 39.2 & 32.7 & 159.88 & $2.56 \times 10^{-22}$ & $4370^{+470}_{-580}$ & $3330^{+2350}_{-1770}$ \\  
        ~ & 54.5 & 19.1 & 158.81 & $1.98 \times 10^{-22}$ & ~ & ~ \\  
        \hline
        ~ & 37.7 & 27.6 & 158.57 & $2.38 \times 10^{-22}$ & ~ & ~ \\  
        \textbf{GW190803\_022701} & 35.2 & 31 & 193.27 & $2.47 \times 10^{-22}$ & $4330^{+470}_{-950}$ & $3190^{+1630}_{-1470}$ \\  
        ~ & 47.5 & 19.1 & 183.88 & $2.00 \times 10^{-22}$ & ~ & ~ \\  
        \hline
        ~ & 46.2 & 30.6 & 174.74 & $1.43 \times 10^{-22}$ & ~ & ~ \\  
        \textbf{GW190805\_211137} & 42.4 & 35 & 179.62 & $1.50 \times 10^{-22}$ & $8790^{+920}_{-980}$ & $6130^{+3720}_{-3080}$ \\  
        ~ & 61.6 & 19.3 & 163.97 & $1.12 \times 10^{-22}$ & ~ & ~ \\  
        \hline
        ~ & 23.3 & 2.6 & 397.19 & $4.52 \times 10^{-22}$ & ~ & ~ \\  
        \textbf{GW190814} & 21.9 & 2.7 & 385.9 & $4.65 \times 10^{-22}$ & $290^{+30}_{-40}$ & $230^{+40}_{-50}$ \\  
        ~ & 24.7 & 2.5 & 382.79 & $4.38 \times 10^{-22}$ & ~ & ~ \\  
        \hline
        ~ & 31.9 & 25.8 & 186.38 & $3.30 \times 10^{-22}$ & ~ & ~ \\  
        \textbf{GW190828\_063405} & 30.7 & 27.8 & 232.45 & $3.37 \times 10^{-22}$ & $2680^{+530}_{-430}$ & $2070^{+650}_{-920}$ \\  
        ~ & 37.3 & 20.5 & 201.6 & $3.03 \times 10^{-22}$ & ~ & ~ \\  
        \hline
        \end{tabular}

\centering
  \footnotesize
\begin{tabular}{@{}llllllll}
    
    \textbf{Event} & $m_1$ $(M_\odot)$ & $m_2$ $(M_\odot)$ & \textbf{$f_{GW}$(Hz)} & \textbf{$h_0$ (max strain)} & \textbf{$d_{Model}$ [Mpc]} & \textbf{$d_{GWTC}$ [Mpc]} \\ \hline
        ~ & 23.7 & 10.4 & 341.47 & $2.21 \times 10^{-22}$ & ~ & ~ \\  
        \textbf{GW190828\_065509} & 17 & 14.2 & 361.27 & $2.40 \times 10^{-22}$ & $1940^{+260}_{-270}$ & $1540^{+690}_{-650}$ \\  
        ~ & 30.5 & 8.2 & 258.65 & $1.93 \times 10^{-22}$ & ~ & ~ \\  
        \hline
        ~ & 43.8 & 34.2 & 155.3 & $4.74 \times 10^{-22}$ & ~ & ~ \\  
        \textbf{GW190910\_112807} & 40.8 & 37 & 140.62 & $6.09 \times 10^{-22}$ & $2060^{+700}_{-390}$ & $1520^{+1090}_{-630}$ \\  
        ~ & 51.4 & 26.9 & 158.34 & $5.49 \times 10^{-22}$ & ~ & ~ \\  
        \hline
        ~ & 32.6 & 24.5 & 191.2 & $3.82 \times 10^{-22}$ & ~ & ~ \\  
        \textbf{GW190915\_235702} & 29.4 & 27.7 & 212.14 & $3.90 \times 10^{-22}$ & $2190^{+340}_{-250}$ & $1750^{+710}_{-650}$ \\  
        ~ & 41.4 & 18.7 & 182.09 & $3.47 \times 10^{-22}$ & ~ & ~ \\  
        \hline
        ~ & 43.8 & 23.3 & 163.07 & $1.46 \times 10^{-22}$ & ~ & ~ \\  
        \textbf{GW190916\_200658} & 35.8 & 31.2 & 191.05 & $1.62 \times 10^{-22}$ & $6140^{+1200}_{-760}$ & $4940^{+3710}_{-2380}$ \\  
        ~ & 63.7 & 13.3 & 130.09 & $1.02 \times 10^{-22}$ & ~ & ~ \\  
        \hline
        ~ & 9.7 & 2.1 & 805.36 & $1.09 \times 10^{-22}$ & ~ & ~ \\  
        \textbf{GW190917\_114630} & 5.8 & 3.2 & 1204.02 & $1.36 \times 10^{-22}$ & $860^{+120}_{-100}$ & $720^{+300}_{-310}$ \\  
        ~ & 13.1 & 1.7 & 630.7 & $9.30 \times 10^{-23}$ & ~ & ~ \\  
        \hline
        ~ & 8.8 & 5.1 & 893.84 & $2.79 \times 10^{-22}$ & ~ & ~ \\  
        \textbf{GW190924\_021846} & 7 & 6.3 & 867.21 & $2.88 \times 10^{-22}$ & $700^{+110}_{-100}$ & $550^{+220}_{-220}$ \\  
        ~ & 13.1 & 3.6 & 612.6 & $2.37 \times 10^{-22}$ & ~ & ~ \\  
        \hline
        ~ & 20.8 & 15.5 & 364.94 & $4.56 \times 10^{-22}$ & ~ & ~ \\  
        \textbf{GW190925\_232845} & 17.9 & 18 & 360.22 & $4.62 \times 10^{-22}$ & $1280^{+150}_{-250}$ & $930^{+460}_{-350}$ \\  
        ~ & 27.3 & 11.9 & 276.77 & $4.19 \times 10^{-22}$ & ~ & ~ \\  
        \hline
        ~ & 41.1 & 20.4 & 160.75 & $1.96 \times 10^{-22}$ & ~ & ~ \\  
        \textbf{GW190926\_050336} & 31.8 & 28.6 & 172.09 & $2.20 \times 10^{-22}$ & $3830^{+450}_{-520}$ & $3280^{+3400}_{-1730}$ \\  
        ~ & 61.9 & 12.2 & 119.08 & $1.41 \times 10^{-22}$ & ~ & ~ \\  
        \hline
       ~ & 66.3 & 26.8 & 129.87 & $2.87 \times 10^{-22}$ & ~ & ~ \\  
        \textbf{GW190929\_012149} & 49.7 & 41.5 & 116.05 & $3.48 \times 10^{-22}$ & $3830^{+740}_{-460}$ & $3130^{+2510}_{-1370}$ \\  
        ~ & 87.9 & 16.2 & 95.86 & $1.98 \times 10^{-22}$ & ~ & ~ \\  
        \hline
        ~ & 14.2 & 6.9 & 601.86 & $2.85 \times 10^{-22}$ & ~ & ~ \\  
        \textbf{GW190930\_133541} & 10.2 & 9.3 & 687.56 & $3.02 \times 10^{-22}$ & $1060^{+130}_{-140}$ & $770^{+320}_{-320}$ \\  
        ~ & 22.2 & 4.8 & 427.92 & $2.34 \times 10^{-22}$ & ~ & ~ \\  
        \hline
 \end{tabular}

\raggedright
\section{GWTC-1}
\centering
\footnotesize
\begin{tabular}{@{}llllllll}
    
    \textbf{Event} & $m_1$ $(M_\odot)$ & $m_2$ $(M_\odot)$ & \textbf{$f_{GW}$(Hz)} & \textbf{$h_0$ (max strain)} & \textbf{$d_{Model}$ [Mpc]} & \textbf{$d_{GWTC}$ [Mpc]} \\ \hline
        ~ & 35.60 & 30.60 & 161.04 & $1.79 \times 10^{-21}$ & ~ & ~ \\  
        \textbf{GW150914} & 40.3 & 26.2 & 186.26 & $1.72 \times 10^{-21}$ & $540^{+100}_{-60}$ & $440^{+150}_{-170}$ \\  
        ~ & 33.6 & 32.5 & 160.78 & $1.79 \times 10^{-21}$ & ~ & ~ \\  
        \hline
        ~ & 23.2 & 13.6 & 298.36 & $3.78 \times 10^{-22}$ & ~ & ~ \\  
        \textbf{GW151012} & 38.1 & 8.8 & 220.24 & $3.03 \times 10^{-22}$ & $1130^{+210}_{-120}$ & $1080^{+550}_{-490}$ \\  
        ~ & 17.7 & 17.7 & 345.92 & $3.92 \times 10^{-22}$ & ~ & ~ \\  
        \hline
        ~ & 13.7 & 7.7 & 561.17 & $5.21 \times 10^{-22}$ & ~ & ~ \\  
        \textbf{GW151226} & 22.5 & 5.2 & 398.78 & $4.31 \times 10^{-22}$ & $590^{+120}_{-70}$ & $450^{+180}_{-190}$ \\  
        ~ & 10.5 & 9.9 & 682.11 & $5.42 \times 10^{-22}$ & ~ & ~ \\  
        \hline
        ~ & 30.8 & 20 & 252.75 & $5.83 \times 10^{-22}$ & ~ & ~ \\  
        \textbf{GW170104} & 38.1 & 15.4 & 212.24 & $5.19 \times 10^{-22}$ & $1240^{+160}_{-140}$ & $990^{+440}_{-430}$ \\  
        ~ & 25.2 & 24.9 & 219.55 & $6.05 \times 10^{-22}$ & ~ & ~ \\  
        \hline
        ~ & 11 & 7.6 & 635.17 & $6.70 \times 10^{-22}$ & ~ & ~ \\  
        \textbf{GW170608} & 16.5 & 5.4 & 560.25 & $5.91 \times 10^{-22}$ & $420^{+60}_{-70}$ & $320^{+120}_{-110}$ \\  
        ~ & 9.3 & 9 & 591.09 & $6.83 \times 10^{-22}$ & ~ & ~ \\  
        \hline
        ~ & 50.2 & 34 & 134.74 & $3.40 \times 10^{-22}$ & ~ & ~ \\  
        \textbf{GW170729} & 66.4 & 23.9 & 140.68 & $2.91 \times 10^{-22}$ & $3760^{+550}_{-540}$ & $2840^{+1400}_{-1360}$ \\  
        ~ & 43.1 & 40 & 146.68 & $3.49 \times 10^{-22}$ & ~ & ~ \\  
        \hline
        ~ & 35 & 23.8 & 194.81 & $6.56 \times 10^{-22}$ & ~ & ~ \\  
        \textbf{GW170809} & 43.3 & 18.6 & 180.98 & $5.94 \times 10^{-22}$ & $1320^{+140}_{-140}$ & $1030^{+320}_{-390}$ \\  
        ~ & 29.1 & 28.9 & 202.24 & $6.74 \times 10^{-22}$ & ~ & ~ \\  
        \hline
        ~ & 30.6 & 25.2 & 206.75 & $1.10 \times 10^{-21}$ & ~ & ~ \\  
        \textbf{GW170814} & 36.2 & 21.2 & 211.31 & $1.06 \times 10^{-21}$ & $780^{+100}_{-90}$ & $600^{+150}_{-220}$ \\  
        ~ & 28 & 27.6 & 211.35 & $1.11 \times 10^{-21}$ & ~ & ~ \\  
        \hline
        ~ & 35.4 & 26.7 & 175.19 & $6.85 \times 10^{-22}$ & ~ & ~ \\  
        \textbf{GW170818} & 42.9 & 21.5 & 171.87 & $6.38 \times 10^{-22}$ & $1320^{+140}_{-150}$ & $1060^{+420}_{-380}$ \\  
        ~ & 31 & 30.7 & 179.73 & $6.96 \times 10^{-22}$ & ~ & ~ \\  
        \hline
        ~ & 39.5 & 29 & 172.48 & $4.11 \times 10^{-22}$ & ~ & ~ \\  
        \textbf{GW170823} & 50.7 & 21.2 & 164.82 & $3.62 \times 10^{-22}$ & $2560^{+320}_{-300}$ & $1940^{+970}_{-900}$ \\  
        ~ & 35.7 & 32.8 & 184.08 & $4.22 \times 10^{-22}$ & ~ & ~ \\  
    \end{tabular}
    \label{GWTC-1}



\section*{}
\raggedright
\textbf{Financial Statement:}
This research received no specific grant from any funding agency in the public, commercial, or not-for-profit sectors.

\textbf{Ethics Declaration:}
Not Applicable.

\bibliography{sn-bibliography}

\end{document}